\documentclass[sigconf, nonacm]{acmart}

\setcopyright{acmlicensed}
\copyrightyear{2018}
\acmYear{2018}
\acmDOI{XXXXXXX.XXXXXXX}
\acmConference[Conference acronym 'XX]{Make sure to enter the correct
  conference title from your rights confirmation email}{June 03--05,
  2018}{Woodstock, NY}
\acmISBN{978-1-4503-XXXX-X/2018/06}

\usepackage{xurl}
\usepackage{subcaption}
\usepackage{enumitem}
\usepackage{graphicx}
\usepackage{tabularx}
\usepackage{caption}
\usepackage{multirow}
\usepackage{array}
\usepackage{pifont}
\usepackage{xcolor}

\usepackage{booktabs,makecell,array}
\newcolumntype{L}[1]{>{\raggedright\arraybackslash}p{#1}}
\newcolumntype{C}[1]{>{\centering\arraybackslash}p{#1}}
\newcolumntype{T}{@{\,\,\,\,}c@{\,\,\(\mid\)\,\,}c@{\,\,\(\mid\)\,\,}c@{\,\,\,\,}}

\usepackage[normalem]{ulem} 

\newcommand{\myline}[1]{{\smallskip\noindent\textbf{#1.}}}

\newcommand{\mynewline}[1]{{\smallskip\noindent\textbf{#1?}}}

\setlist{
  listparindent=\parindent,
  parsep=0pt,
}

\usepackage{subcaption}
\begin{document}
\title{O$^3$-LSM: Maximizing Disaggregated LSM Write Performance via Three-Layer Offloading}

\author{Qi Lin}
\affiliation{
  \institution{Arizona State University}
  \country{USA}
}
\email{qlin36@asu.edu}

\author{Gangqi Huang}
\affiliation{
  \institution{Arizona State University}
  \country{USA}
}
\email{alexhuang1403@gmail.com}

\author{Te Guo}
\affiliation{
  \institution{Purdue University}
  \country{USA}
}
\email{guo777@purdue.edu}

\author{Chang Guo}
\affiliation{
  \institution{Arizona State University}
  \country{USA}
}
\email{cguo51@asu.edu}

\author{Viraj Thakkar}
\affiliation{
  \institution{Arizona State University}
  \country{USA}
}
\email{viraj.dt@asu.edu}

\author{Zichen Zhu}
\affiliation{
  \institution{Google}
  \country{USA}
}
\email{zczhu@bu.edu}

\author{Jianguo Wang}
\affiliation{
  \institution{Purdue University}
  \country{USA}
}
\email{csjgwang@purdue.edu}

\author{Zhichao Cao}
\affiliation{
  \institution{Arizona State University}
  \country{USA}
}
\email{zhichao.cao@asu.edu}
\begin{abstract}
Log-Structured Merge-tree-based Key-Value Stores (LSM-KVS) have been optimized and redesigned for disaggregated storage via techniques such as compaction offloading to reduce the network I/Os between compute and storage.
However, the constrained memory space and slow flush at the compute node severely limit the overall write throughput of existing optimizations.
In this paper, we propose \textbf{O$^3$-LSM}, a fundamental new LSM-KVS architecture, that leverages the shared Disaggregated Memory (DM) to support a \textit{three-layer} offloading, i.e., memtable \textbf{O}ffloading, flush \textbf{O}ffloading, and the existing compaction \textbf{O}ffloading. Compared to the existing disaggregated LSM-KVS with compaction offloading only, O$^3$-LSM maximizes the write performance by addressing the above issues.

O$^3$-LSM first leverages a novel \textit{DM-Optimized Memtable} to achieve \textit{dynamic memtable offloading}, which extends the write buffer while enabling fast, asynchronous, and parallel memtable transmission.
Second, we propose \textit{Collaborative Flush Offloading} that decouples the flush control plane from execution and supports memtable flush offloading at any node with dedicated scheduling and global optimizations.
Third, O$^3$-LSM is further improved with the \textit{Shard-Level Optimization}, which partitions the memtable into shards based on disjoint key-ranges that can be transferred and flushed independently, unlocking parallelism across shards. 
Besides, to mitigate slow lookups in the disaggregated setting, O$^3$-LSM also employs an adaptive \textit{Cache-Enhanced Read Delegation} mechanism to combine a compact local cache with DM-assisted memtable delegated read. 
Our evaluation shows that O$^3$-LSM achieves up to \textbf{4.5$X$} write, \textbf{5.2$X$} range query, and \textbf{1.8$X$} point lookup throughput improvement, and up to \textbf{76\%} P99 latency reduction compared with Disaggregated-RocksDB, CaaS-LSM, and Nova-LSM.
\end{abstract}

\settopmatter{printfolios=true}
\maketitle

\section{Introduction}

With the fast development of cloud computing and high-speed networks, Disaggregated Data Centers (\textbf{DDCs}) have been widely adopted in the industry to address the challenges of resource utilization, load-balancing, and scalability in traditional monolithic setups \cite{lin2020ddc, zhang2020ddc, zhang2020rethinking, wang2023disaggregated}. 
DDCs achieves high flexibility by decoupling compute, memory, and storage into independently scalable resource pools, such as compute clusters, Disaggregated Storage (\textbf{DS}), and Disaggregated Memory (\textbf{DM}). 
Driven by this architectural evolution, \textbf{Log-Structured Merge-tree-based Key–Value Stores (LSM-KVS)} are being redesigned for DDCs, resulting in disaggregated LSM-KVS, to fully exploit the DDC model for maximizing both resource utilization and performance.

In a typical disaggregated LSM-KVS, ingested Key-Value pairs (KV-pairs) are initially batched into the write buffer as memtables within one or more compute nodes (\textbf{CNs}). 
When a memtable becomes full, it will be marked as immutable and flushed from the CN to DS as a Sorted String Table (SST) file through the network. 
To ensure better space efficiency and read performance, SST files in DS will be periodically read back to CN through the network, merged into new SST files, and written back to DS.
The periodic merging, also termed \textit{compaction}, runs in the background process, and is designed not to interfere with the foreground user ingestion process.
The decoupling provided by DDCs is a natural fit for this task, as it allows \textbf{compaction processes to be separated from CNs and offloaded into DS}, significantly reducing resource contention and network traffic on CNs. 
As such, many state-of-the-art disaggregated LSM-KVS -- including Disaggregated-RocksDB at Meta \cite{dong2023rocksdb}, TerarkDB at Bytedance \cite{terark}, IS-HBase \cite{cao2022is-hbase}, Hailstorm \cite{Bindschaedler2020hailstorm}, Nova-LSM \cite{Huang_2021}, Rocks-Cloud \cite{rockscloud}, and CaaS-LSM \cite{yu2024caas-lsm} -- adopt this design by offloading compaction to DS nodes or redistributing it across CNs to mitigate write performance penalties caused by data movement between CN and DS during compaction.

However, compaction offloading does not effectively mitigate \textit{write slowdowns and stalls} caused by \textbf{limited write-buffer memory and slow flush operations}.
When the write-buffer limit is reached during data ingestion, new writes are typically throttled or blocked until the memory space is freed via flushing, which directly cause slowdowns and stalls.
Given that an LSM-KVS instance usually requires substantial memory on both the read path (i.e., block cache) and the write path (i.e., memtables) \cite{cao2020characterizing},
the memory of CNs is often limited, particularly in DDCs setups where a single CN can simultaneously host tens of LSM-KVS instances and other applications .
In addition to limited memory, memtable flush also incurs heavy network I/O to DS with high latency \cite{dong2023rocksdb,yu2024caas-lsm,cao2022is-hbase}.
This network traffic contends with other background and foreground I/Os (e.g., WAL appends/syncs, compaction, and SST file reads) on the CN, which prolongs the flush process and leads to further write slowdowns and stalls \cite{yu2023adoc}. 
Moreover, flush operations can be also \textit{suspended} if the number of SST files at level 0 ($L_0$) reaches the predefined limit due to serial and slow $L_0$ compactions \cite{balmau2019silk, yu2023adoc}. 

With the widely deployed remote-memory technologies such as RDMA \cite{inifiniband}, DDCs are able to provide the shared Disaggregated Memory (DM) pools for CN as the secondary tier of memory. 
Given this capability, a straightforward solution to constrained write buffer is to extend the disaggregated LSM-KVS write buffer by offloading memtables to DM as a short-term staging area rather than flushing immediately to DS. 
However, \textbf{simply offloading memtables to DM instead leads to lower ingestion throughput and higher read latency}.


\noindent \textbf{Problem 1: Costly remote memtable shipping and rebuild.}
Even with fast RDMA, accessing remote memory is slower than local DRAM, so moving a memtable to DM is not a cheap memcpy. The {\color{black}memtable} traverses the RNIC, PCIe, and fabric, adding non-negligible transfer latency and bandwidth consumption. Moreover, a memtable is {\color{black}pointer-intensive} (e.g., skiplist nodes allocated from arenas), so a raw byte copy invalidates addresses on the DM. Making it searchable on DM requires extra updating of pointers and reconstruction of the index. Therefore, both memtable transfer and memtable rebuild at DM will have explicit performance penalties.

\noindent \textbf{Problem 2: Inefficient memtable flush.}
Offloading immutable memtables to DM is only the first step. For disaggregated LSM-KVS, memtables still need to be flushed to DS for persistence. However, DM cannot directly flush these offloaded memtables to DS, since flush requires KV-pair sorting and merging, compression, SST file construction, and Manifest updates, which are missing in DM. Therefore, the memtables need to be read back to the CN and perform the flush locally by the owning LSM-KVS. This process adds an extra network hop, reintroduces serialization and deserialization overhead, and may be further hindered by I/O contention with other foreground and background operations on the same CN. Meanwhile, as many DM-resident memtables are flushed to DS, $L_0$ SST files accumulate and exacerbate the issue of slow $L_0$ compaction. All of these effects degrade overall performance.

\noindent \textbf{Problem 3: Slower memtable search at DM.}
After offloading memtables to DM, searching them can incur significant read performance regression. On one hand, a key lookup within a single memtable requires multiple memory accesses due to its semi-sorted, {\color{black}pointer-intensive} structure (e.g., skiplist or tree). When the memtable is at DM, this process will involve multiple remote memory access round-trips and thus lead to higher lookup latency. On the other hand, a read query may need to search all the accumulated memtables belonging to this LSM-KVS at DM. With more memtables accumulated at DM, the read performance can be worse.

To address the aforementioned challenges of memtable offloading to DM, we propose \textbf{O$^3$-LSM}, a fundamental new LSM-KVS architecture that leverages the shared Disaggregated Memory (RDMA-Based) to support a \textit{three-layer} offloading, encompassing memtable \textbf{O}ffloading, flush \textbf{O}ffloading, and the existing compaction \textbf{O}ffloading. Compared to the state-of-the-art disaggregated LSM-KVS with compaction offloading only, O$^3$-LSM achieves superior performance by introducing the following four key innovations:

\begin{itemize}[leftmargin=*, topsep=2pt]
\item \textbf{DM-Optimized Memtable}: \textcolor{black}{O$^3$-LSM introduces a DM-optimized memtable that is designed to be directly searchable on DM without heavy reconstruction. Instead of transferring and rebuilding {\color{black}pointer-intensive} memtables remotely at DM, O$^3$-LSM writes memtables in a DM-friendly, memory-contiguous layout with index-data separation so that offloading becomes a cheap data transfer and the memtable can be queried on DM as is.}


\item \textbf{Collaborative Flush Offloading}: O$^3$-LSM develops a lightweight, judiciously designed multi-phase offloading protocol, which allows a memtable flush to be executed on \textit{any} node with DM access. 
This protocol supports multiple execution modes (\textbf{Local}, \textbf{In-DM}, \textbf{Remote-CN}) and various controlling messages, providing the foundation for a \textbf{contention-aware flush scheduler}.
This scheduler coordinates all flush operations by dynamically adapting to workload, data locality, and resource utilization and also by handling potential failures to maximize overall write performance with ensuring atomicity and isolation properties.

\item \textbf{Asynchronous Sharding}: 
O$^3$-LSM also partitions each memtable into a set of non-overlapping key-range shards by dividing the KV-block into multiple ones, enabling asynchronous, highly parallel transfer to DM.
When flushing multiple memtables,
O$^3$-LSM aggregates KV-shard blocks of the same key range across different memtables.
This approach produces non-overlapping, shard-aligned $L_0$ SST files, effectively breaking down a serial $L_0$ compaction into multiple parallel tasks, and thereby significantly reducing overall write stalls.
\item \textbf{Cache-Enhanced Read Delegation}: O$^3$-LSM accelerates reads by the adaptive read delegation. 
Leveraging the DM-optimized memtable structure, O$^3$-LSM maintains a small key-offset cache at CN to decide the read path on a per-read basis.
This adaptive mechanism determines whether to retrieve data directly from DM using one-sided \texttt{RDMA\_READ} or delegate the read to DM with two-sided \texttt{RDMA\_SEND}, reducing unnecessary remote network traversals and mitigating extra CN memory usage.
\end{itemize}

\begin{figure*}[!t]
  \begin{minipage}[b]{0.48\textwidth}
    \centering
    \includegraphics[width=\textwidth]{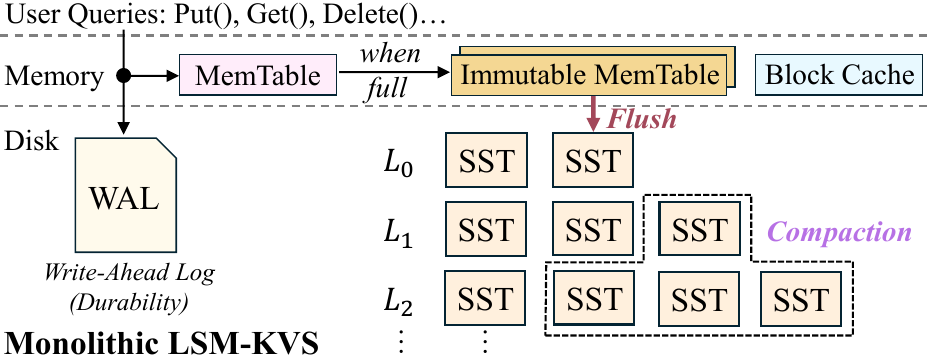}
    \subcaption{Monolithic LSM-KVS Architecture.}
    \label{fig:kvs}
  \end{minipage}
   \hfill
  \begin{minipage}[b]{0.48\textwidth}
    \centering
    \includegraphics[width=\textwidth]{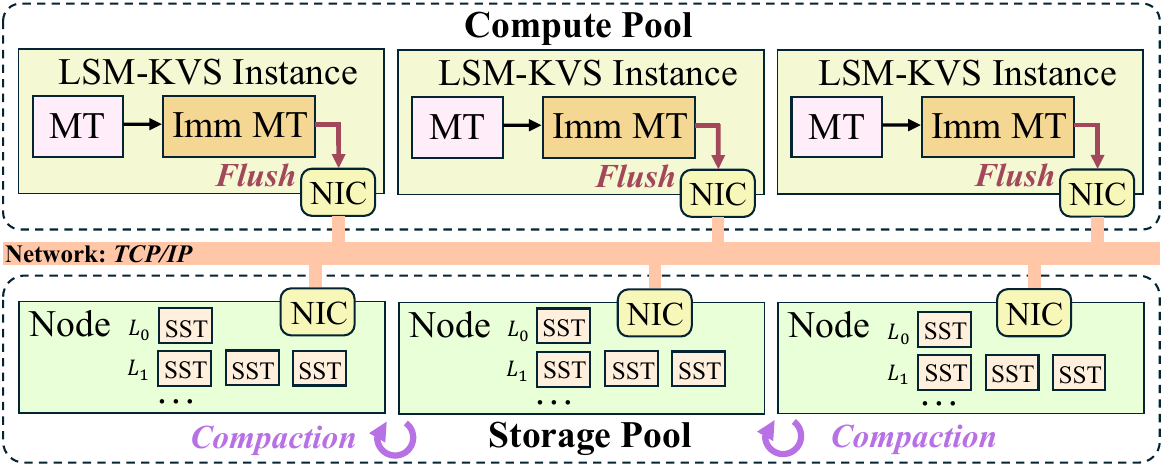}
    \subcaption{\color{black}Disaggregated LSM-KVS Architecture.}
    \label{fig:dis-kvs}
  \end{minipage}
  \caption{Comparison between Monolithic and Disaggregated LSM-KVS Architectures.}
    \label{fig:background_architecture}
\end{figure*}

\myline{Implementation and Overall Evaluation Results} We prototyped O$^3$-LSM on RocksDB v8.2.0 with 30,000 lines of code change, open-sourced on \textbf{Github}\footnote{\textcolor{black}{https://github.com/argetterxx/O3-LSM}}. 
Evaluations on CloudLab \cite{cloudlab} against state-of-the-art LSM-KVS schemes, including Disagg-RocksDB \cite{dong2023rocksdb}, CaaS-LSM \cite{yu2024caas-lsm}, and Nova-LSM \cite{Huang_2021} under identical write buffer limits. O$^3$-LSM demonstrates performance gains in both throughput and P99 latency across a variety of workloads. For random write workload, O$^3$-LSM achieves \textbf{4.5$X$}, \textbf{3.4$X$}, and \textbf{4.4$X$} throughput improvements while slashing P99 latency by up to \textbf{60\%}, outperforming Disagg-RocksDB, CaaS-LSM, and Nova-LSM. For random read workload, O$^3$-LSM delivers \textbf{1.8$X$}, \textbf{0.6$X$}, and \textbf{0.3$X$} higher throughput with a P99 latency reduction of up to \textbf{69\%} compared to these baselines. For range query workload, O$^3$-LSM achieves up to \textbf{5.2$X$} throughput improvements while reducing P99 latency by up to \textbf{22\%}. Even in mixed workloads (50\% read and 50\% write operations), O$^3$-LSM also exhibits up to \textbf{3$X$}, \textbf{2.3$X$}, and \textbf{1.9$X$} throughput improvements while reducing P99 latency by up to \textbf{76\%}. 
For the real-world application, Kvrocks~\cite{apache_kvrocks}, O$^3$-LSM achieves up to \textbf{3.4$X$} throughput improvements while reducing P99 latency by up to \textbf{54\%}.

\section{Background}
\subsection{LSM-based Key-Value Stores}
LSM-KVS like RocksDB \cite{rocksdb} and LevelDB \cite{leveldb} are optimized for high-performance writes via log-structured storage. As shown in Figure~\ref{fig:kvs}, LSM-KVS appends KV-pairs to a Write-Ahead-Log (WAL) and an in-memory active memtable. Full active memtables become immutable and are serialized as Sorted String Table (SST) files through a flush process into $L_0$. Consequently, each $L_0$ SST file may contain KV-pairs spanning the entire key space. SST files are organized across multiple levels. $L_0$ files are compacted sequentially into $L_1$ because their key ranges often overlap. In contrast, levels $L_n$ ($n \geqslant 1$) have disjoint ranges, allowing parallel compaction with overlapping files in $L_{n+1}$. To manage backpressure, LSM-KVS triggers write slowdowns or stalls if immutable memtables or $L_0$ files become excessive. For read queries, LSM-KVS maintains a block cache in memory to cache the data/metadata blocks from SST files. 

\subsection{Disaggregated LSM-based Key-Value Stores}
Disaggregated Data Centers (DDCs) have gained traction as a way to improve resource utilization and elasticity by decoupling compute, memory, and storage into independent resource pools interconnected by high-speed fabrics \cite{zhang2020ddc,angel2020disaggregation,dong2023rocksdb}. Optimizing LSM-KVS for DDCs (called disaggregated LSM-KVS), especially stores the persistent data (e.g., SST files) at Disaggregated Storage (e.g., Tectonic at Meta \cite{tectonic}, GFS at Google \cite{ghemawat2003google}, and HDFS \cite{hdfs}) become popular \cite{yu2024caas-lsm, dong2023rocksdb}. As shown in Figure~\ref{fig:dis-kvs}, a typical disaggregated LSM-KVS design keeps memtables on compute nodes (CNs) and stores SSTables on DS nodes (DSNs), with CNs accessing DSNs via file or block interfaces. This separation enables better load balancing while retaining comparable write performance to local deployments \cite{dong2023rocksdb}. 
Recent work has optimized disaggregated LSM-KVS over DS, primarily focusing on compaction offloading \cite{dong2023rocksdb, terark, Bindschaedler2020hailstorm, yu2024caas-lsm}.

However, bottlenecks remain regarding CN memory. LSM-KVS requires substantial memory for both the read path (block cache) and the write path (memtable buffers) \cite{cao2020characterizing}. CNs often host tens of LSM-KVS instances, making it infeasible to allocate large memory to each instance \cite{zippydb}. Insufficient memory at LSM-KVS instances triggers more frequent flush operations to persist memtables quickly. These flushes amplify network I/Os to DS, where higher latency and shared bandwidth prolong the process and can cause write slowdowns or stalls. The problem is exacerbated by $L_0$ compaction, where $L_0$ files span the full key space. Therefore, $L_0$ files can only be compacted sequentially, and high writing pressure can easily hit the $L_0$ file cap, leading to throttle or block flushes \cite{balmau2019silk,yu2023adoc}. Consequently, memory pressure and slow flushes remain core bottlenecks in current disaggregated LSM-KVS designs.

\begin{figure}[t]
  \centering
  \begin{minipage}[t]{0.49\textwidth}
    \centering
    \includegraphics[width=\linewidth]{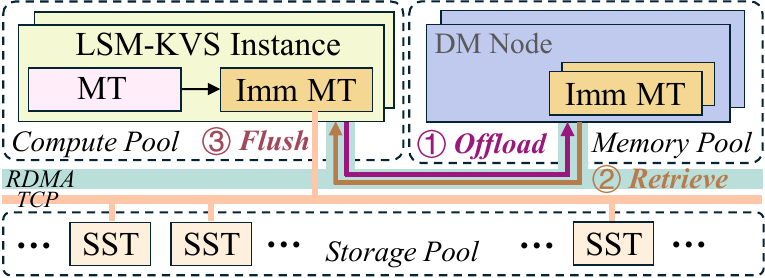}
    \captionsetup{labelformat=parens, labelsep=space}
  \end{minipage}
  \caption{A naive solution that offloads and retrieves immutable memtables via the Memory Pool.}
  \label{fig:easy-solution}
\end{figure}

\subsection{Opportunities with Disaggregated Memory: A New Tier for Memtables}
RDMA \cite{hpcc12rdma} and CXL \cite{cxl} have enabled disaggregated memory (DM) in DDCs, providing memory pooling and sharing. Unlike DS, DM is typically non-persistent to maximize performance. Researchers have explored in-memory LSM-KVS on DM, such as dLSM \cite{wang2023dlsm}, which stores SSTables on DM and caches memtables in CNs. While dLSM lacks persistence guarantees and differs from Disaggregated-RocksDB or CaaS-LSM, DM remains a promising substrate to alleviate write-path memory pressure and enhance performance in persistent, DS-based systems.


Beyond write-path opportunities, DM differs from local memory in read-path behaviors. RDMA-based DM typically uses two methods. One-sided \texttt{RDMA\_READ} allows compute nodes to pull bytes directly using remote addresses and rkeys, bypassing the remote CPU but potentially incurring extra round trips. Alternatively, two-sided \texttt{RDMA\_SEND}/\texttt{RDMA\_RECV} involves a remote thread reading local DRAM to reply in one exchange, which reduces round trips but consumes remote CPU and adds RPC overhead. In practice, one-sided reads prioritize zero CPU usage over network latency, whereas two-sided reads minimize round trips at the cost of remote CPU.

\begin{table}[t]
\centering
\setlength{\tabcolsep}{3pt}           
\renewcommand{\arraystretch}{0.95}    
\small
\begin{tabular}{@{}lccccc@{}}
\toprule
Metric & K=2 & K=4 & K=8 & K=16 & K=32 \\
\midrule
Memtable-induced write stalls (count)  & 737 & 614 & 268 & 115 & 34 \\
Share of total stalls (\%)         & 96  & 85  & 60  & 30  & 11 \\
Write-stall latency P99 (ms)    & 2   & 6   & 10  & 14  & 18 \\
\bottomrule
\end{tabular}
\caption{Impact of DM-offloaded memtables on write stalls}
\label{tab:write-stall}
\end{table}

\begin{figure*}[!t]
  \centering
  \includegraphics[width=1.0\linewidth]{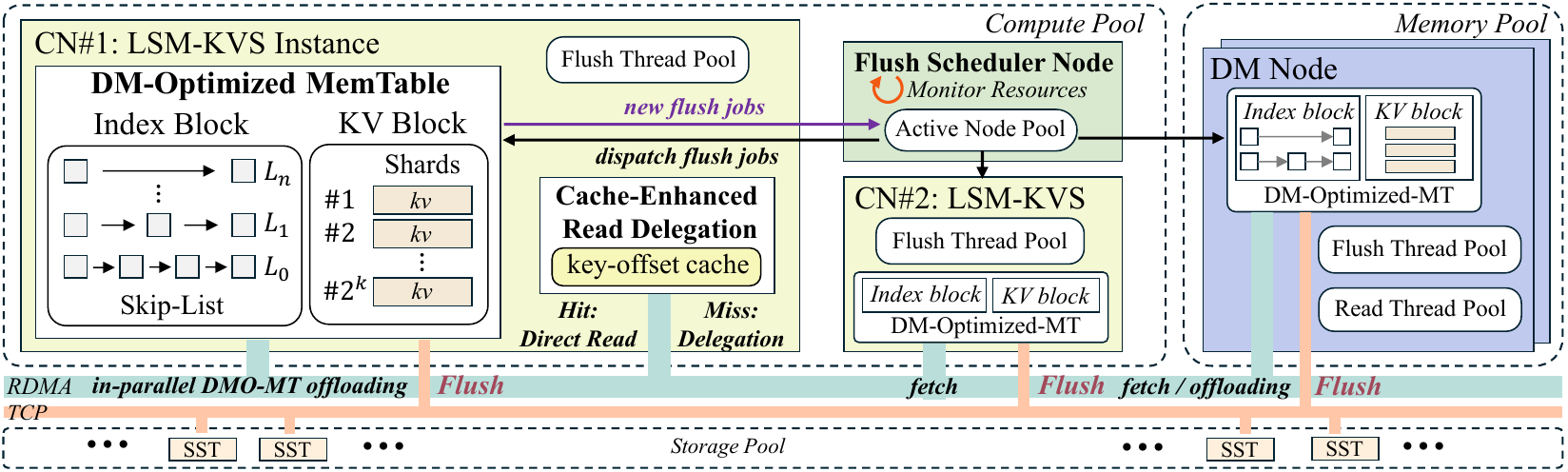}
  \caption{The overall architecture of O$^3$-LSM.}
  \label{fig:figure1}
  \vspace{-5pt}
\end{figure*}

\section{Motivations and Challenges}
Disaggregated storage (DS) offers elastic storage capacity for persistent LSM-KVS, but compute-node memory remains a bottleneck, especially with many co-located instances. Disaggregated memory (DM) can act as a low-latency, high-bandwidth tier between compute-node memory and DS. Therefore, leveraging a DM tier to cache additional memtables, while retaining durability of SST files on DS, is a promising way to relieve per-instance memory pressure for high performance without sacrificing persistence.

\subsection{A Naive Solution} \label{sec:naive}
A straightforward approach to relieve local memory pressure at compute nodes (CNs) is to offload immutable memtables to DM, making DM a fast, scalable extension of CN memory (\autoref{fig:easy-solution}). We evaluated this by modifying Disaggregated RocksDB \cite{dong2023rocksdb} to transfer memtables via RDMA while keeping WAL locally and SSTs on HDFS \cite{hdfs}. Using db\_bench "randomwrite" with 116B KV-pairs, we increased the remote memtable limit ($K$) from 2 to 32. As shown in \autoref{tab:write-stall}, this reduced memtable-induced write stalls by 95\%.

However, throughput dropped from 119 to 87 kop/s and p99 write-stall latency rose from 2 to 18 ms. Analysis shows that while stalls decreased, transferring and rebuilding memtables on DM consumes 18.7\% of total time. Furthermore, retrieving memtables from DM for DS flushes adds 7.8\% overhead compared to local flushes. Read performance also suffers, as DM reads take 29 us compared to 4.7 us locally. This latency is dominated by RDMA overhead, including 12 us for \texttt{RDMA\_SEND} and 11 us for \texttt{RDMA\_RECV}. Remote searches also incur extra copying costs. These results indicate that transfer, rebuild, and retrieval overheads can outweigh the benefits of reduced write stalls.

\subsection{Issues and Challenges}
As discussed in Section \ref{sec:naive}, offloading memtables from CN local memory to DM may be able to further improve the performance of disaggregated LSM-KVS. However, addressing the performance issues, including 1) slower write caused by memtable transferring and rebuilding latency, 2) delayed flush due to the memtable retrieval from DM, and 3) read performance regression resulting from searching memtables at DM, is challenging.

\mynewline{How to efficiently transfer and rebuild memtables from CN to DM} Since the remote memory access latency can potentially be multiple times higher than local DRAM access \cite{gao2016network,shoal2019,amaro2020fastswap}, transferring memtables from CN to DM can lead to high latency and may block subsequent memtable writes. The situation can be even worse under highly write-intensive workloads. Additionally, due to the memtable data structure design (e.g., widely used skiplist-based memtable), the memtable consists of a large number of pointers. When the memtable is transferred to DM, all the pointers become invalid and require re-allocation at DM, which can be slow.

\mynewline{How to speed up the memtable flush from DM to DS} 
When immutable memtables are offloaded to DM, the process of flushing them can become even slower when executed with the original flush logic. Directly creating SST files from memtable at DM is difficult, which involves flushing in-memory data to persistent storage while performing key sorting, value serialization, block and index building, compression, metadata management, WAL handling, and snapshot control. Since DM itself does not have flush execution logic, LSM-KVS must read the memtables back from DM to CN to flush the KV-pairs. This read-back process can introduce additional latency and may be further hindered by I/O contentions with other foreground and background I/O operations on the same CN. Furthermore, when a large number of accumulated memtables at DM are flushed to DS, the issue of slow $L_0$ compaction (a well-known problem in LSM-KVS due to the need to handle the entire key-space coverage of each $L_0$ SST file \cite{yu2023adoc}) is exacerbated due to the $L_0$ SST file accumulation. This can result in increased write slowdown or write stall and can significantly impact overall performance.

\mynewline{How to mitigate read performance regression caused by slow memtable search at DM}
All the memtables are accessed during each read query due to the key-range overlaps. When one key is searched in a memtable (e.g., widely used skiplist-based memtable), multiple memory accesses are needed to finally pinpoint the KV-pair (or confirm {\color{black}NotFound}). When memtables are maintained in DM, due to the relatively slower memory access via RDMA, searching one memtable can cause explicitly high latency. Even worse, one read query needs to search multiple memtables until the KV-pair is found (or {\color{black}NotFound} in all memtables), which further amplifies the high read latency.



\section{O$^3$-LSM} \label{sec:design}

\subsection{Architecture Overview}

\autoref{fig:figure1} shows the O$^3$-LSM architecture. Like other disaggregated LSM-KVS\cite{dong2023rocksdb, yu2024caas-lsm}, O$^3$-LSM instances run on CNs and store SST files on DS. It uses a two-layered write buffer integrating local CN DRAM with high-capacity DM. New active memtables are first cached in local memory for fast insertions. They are then offloaded to DM as immutable memtables before the final flush to DS.


First, to tackle the performance challenge of transferring and rebuilding memtables in DM, O$^3$-LSM introduces a novel \textbf{DM-Optimized MemTable} (detailed in Section \ref{sec:skiplist-design}) at both CNs and DM nodes. The DM-Optimized MemTable redesigns the skiplist structure with a pointer-based index-block (containing the linked list nodes of the skiplist) and the one KV-block (KV-block will be further divided into multiple KV-shard blocks if shard-level optimization is applied, as in Section \ref{sec:shard} for more details). The KV-block stores KV-pairs sequentially, and KV-pairs are referenced by their offsets within the KV-block stores by the skiplist index nodes in the index-block. Therefore, KV-block can be directly transferred to DM without pointer reconstruction.

Second, to speed up memtable flushes, we introduce \textbf{Collaborative Flush Offloading} (Section~\ref{sec:remote-flush}), which decouples the flush path from the LSM-KVS instance and makes the flush job remotely executable. We propose a flush offloading protocol to transfer the metadata between the memtable owner and the flush executor. Since DM is shared by all CNs, rather than reading immutable memtables from DM back to the owning CN for flush, memtables at DM will be collaboratively flushed by any selected CN or DM nodes. We introduce a collaborative flush scheduler, which monitors resource information (i.e., I/O bandwidth, CPU utilization) and collects flush job metadata (i.e., memtable offset at DM nodes, original CN). The scheduler will select one CN or DM node with spare CPU and I/O to execute the memtable flush by using the flush offloading protocol, improving resource utilization and speeding up flush.

Third, to further improve scalability and parallelism, as well as $L_0$ compaction, we propose the \textbf{Shard-Level Optimizations}, which partition one memtable table into multiple shards based on the pre-defined key-ranges (the same key-range partition will be applied to all memtables). Thus, one KV-block is partitioned into multiple KV-shard blocks. In this way, multiple KV-shard blocks of one memtable can be transferred asynchronously with high parallelism, which further improves the memtable transferring efficiency. At the same time, instead of flushing each memtable separately, we propose to flush the same key-range from all memtables (i.e., merging all the KV-shard blocks of the same key-range) belonging to the same LSM-KVS at the DM together. The shard-based flush actually combines flush and $L_0$ compaction in a fine-grained way, which significantly reduces the number of SST files in $L_0$ with key-range overlap. O$^3$-LSM thus breaks down flush offloading from coarse-grained, memtable-level tasks to fine-grained, shard-level tasks, enabling greater scalability and parallelism, avoiding the $L_0$ penalties, and significantly improving overall performance.


Finally, to mitigate the high overhead associated with searching immutable memtables in DM through multiple RDMA accesses, O$^3$-LSM employs a \textbf{Cache-Enhanced Read Delegation} (detailed in Section \ref{sec:read delegation}). Hot KV-pairs in memtables at DM are cached in a small key-offset cache at CN (caching the key and its DM offset), and we use one-sided \texttt{RDMA\_READ} to fetch the corresponding KV-pair. If cache misses, O$^3$-LSM initiates the read delegation request to DM via two-sided \texttt{RDMA\_SEND}, where a DM-side CPU (read thread) searches the memtables to significantly reduce round trips. In addition, we incorporate well-known RDMA optimizations (e.g., doorbell batching and inlining) to further improve efficiency.

\textcolor{black}{While $O^3$-LSM is evaluated on RDMA due to its current maturity and availability~\cite{dm,cao2021polardb,li2022dm,wang2023dlsm}, its design is interconnect-agnostic. Even with the cache-coherent, low-latency access provided by emerging technologies like CXL \cite{nag2023cxl, cxl-3.0}, the core principles of $O^3$-LSM remain essential. Specifically, a DM-friendly layout is still essential to mitigate costly remote pointer chasing, and our shard-level offloading and read delegation provide critical orchestration layers to balance flush-intensive workloads and maximize throughput. Thus, $O^3$-LSM remains both applicable and beneficial in a CXL-based disaggregated infrastructure.}

\subsection{DM-Optimized MemTable}
\label{sec:skiplist-design}

Similar to most LSM-KVS implementations \cite{rocksdb, leveldb, wang2023dlsm, terark, dong2023rocksdb, yu2024caas-lsm}, O$^3$-LSM uses skiplist as the data structure to implement the memtable due to its efficient concurrency, fast point lookup, and range queries. A standard skiplist node stores $k$ pointers followed by a KV-pair in contiguous memory. These pointers form $k$ index levels, where higher-level pointers skip nodes to create a sparse index. This structure typically enables $\mathcal{O}(\log n)$ query complexity.

\begin{figure}
  \centering
  \includegraphics[width=1\linewidth]{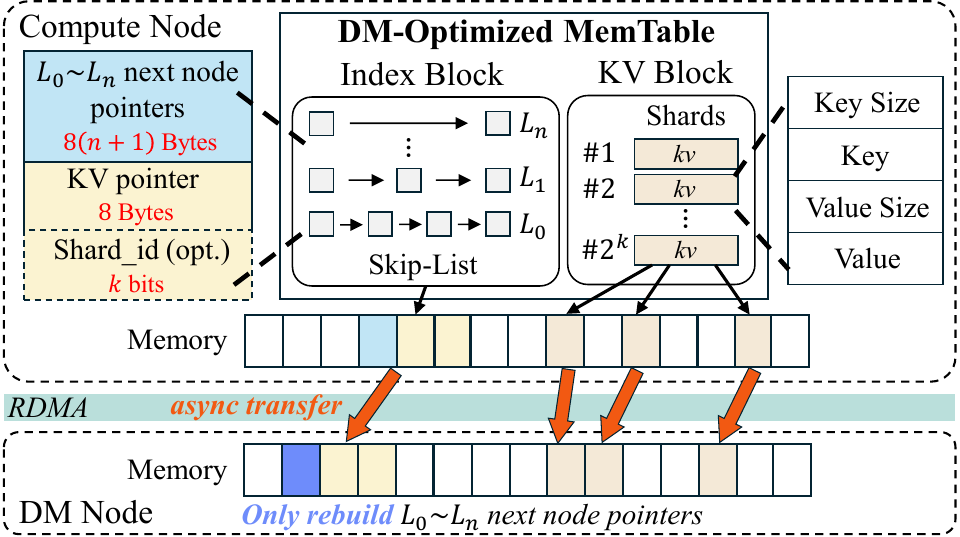}
  \caption{DM-Optimized SkipList-Based MemTable}
  \label{fig:sec4_fig1}
\end{figure}

However, transferring and rebuilding the current skip-list-based memtable to DM are slow and inefficient. First, transferring the entire memtable to DM requires traversing all skiplist nodes at CN and reconstructing them with pointers in DM, which can slow down the memtable offloading process. Second, during the rebuilding process, we need to dynamically allocate the DM memory space for skiplist nodes. Thus, those nodes are scattered in the whole DM memory space, which incurs both space and performance overhead. A straightforward solution is to preallocate a sufficiently large contiguous memory block in DM (e.g., 64MB) for dynamic node allocation in one memtable. However, this strategy can lead to memory fragmentation and space wasting.


To address these, we proposed a \textbf{DM-Optimized MemTable}, as shown in \autoref{fig:sec4_fig1}. We separate the memtable into a pointer-based index-block containing the linked list nodes and one KV-block that stores all the KV-pairs without pointers. Each node in the index block holds: 1) $8(n+1)$ bytes for the next node pointers, 2) a \texttt{KV pointer} with 8 bytes pointing to the corresponding KV-pair offset in the KV-block, and 3) a \texttt{shard\_id} with $k$ bits (if shard-level optimization is applied). In KV-block, KV-pairs are appended sequentially in a contiguous memory block and organized as tuples (\textless key, value, offset\textgreater). This separation allows the memtable to be stored in different, smaller contiguous memory blocks. The KV-block can be directly transferred and stored at a pre-allocated contiguous memory block in DM without further processing. Reconstruction is only needed for the index-block to correct the pointers with the new memory addresses of KV-pairs.

\begin{figure*}
  \centering
    \includegraphics[width=1\linewidth]{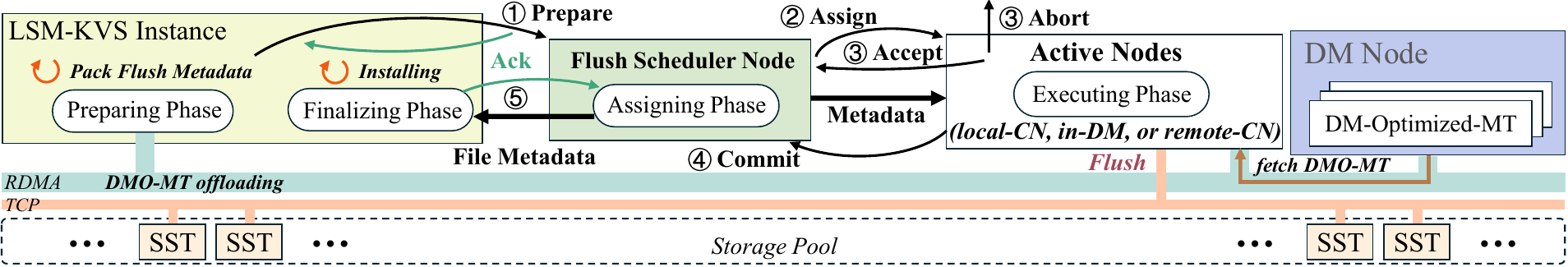}
  \caption{Collaborative Flush offloading}
  \label{fig:remoteflush}
  \vspace{-5pt}
\end{figure*}


When a memtable becomes immutable, we enqueue it into an asynchronous transfer queue. A background worker first serializes the index-block into a compact, contiguous buffer, registers this buffer as a single RDMA memory region, and issues a one-sided \texttt{RDMA\_WRITE} to a pre-allocated region on DM. KV-pairs do not need to be serialized since they are length-prefix encoded in the KV-block as strings. The worker registers the KV-block and transfers it to its designated DM region with one-sided \texttt{RDMA\_WRITE}. 
{\color{black}During the transfer, we maintain a small metadata record (114 bytes in total) containing the start addresses of the index-block and KV-block on both the compute node and DM. This compact record ensures the metadata overhead remains below 0.0002\% for a 64 MB memtable.} After the writes are complete, the DM side reconstructs the index-block by refreshing the pointers of the skiplist nodes. The memtable is then marked searchable on DM, and the transfer job completes. 

Note that the offset information of each KV-pair stored in the index-block does not need to be updated at DM. O$^3$-LSM will correct these fields in the index by adding the appropriate memory offset of the KV-block at query time. Also, KV-block does not need to be updated/reconstructed at DM since all KV-pairs are appended sequentially with length-prefix encoded. For example, suppose the starting memory address of one KV-block on the CN is $A$. After the KV-block is transferred to DM, the starting memory address is $B$. The offset of one KV-pair in the KV-block is $C$, which is stored in one index node at the index-block. Therefore, the correct DM address of the KV-pair can be determined by adding the offset within the KV-block to the offset of the KV-block at DM address (i.e., $C + B - A$), which can be corrected during memtable search in time complexity of $\mathcal{O}(1)$.

\subsection{Collaborative Flush offloading} \label{sec:remote-flush}

In traditional LSM-KVS designs, the flush process is coupled to the LSM-KVS instance that created the flush job. This coupling causes two drawbacks when DM is used as an external write buffer: (1) since DM lacks flush execution logic (e.g., key sorting and flush-metadata management), the initiating LSM-KVS must fetch immutable memtables back from DM and flush them to DS, incurring transfer and (de)serialization overheads; and (2) memtable flush might still {\color{black}contention} for network I/O with foreground reads and background compactions. As DM accumulates memtables, these costs grow and may throttle foreground reads/writes.


To address these challenges, we propose a novel \textbf{Collaborative Flush Offloading} (Figure~\ref{fig:remoteflush}). We decouple the flush control plane from the LSM-KVS instance and make it remotely executable on any node with DM access. The design has two components: 1) a lightweight flush-offloading protocol that prepares, coordinates, and conveys all required metadata, orchestrates workflow and failure handling, and finalizes job state, and 2) a logically centralized flush scheduler that assigns flush jobs to CNs or DM nodes based on load and locality. As shown in Figure~\ref{fig:remoteflush}, O$^3$-LSM offloads flush in four phases, supports three execution modes (local, in-DM, and remote-CN), and uses six control messages.

\myline{Phase 1: \color{black}Preparing} When a memtable flush is invoked, the owning LSM-KVS snapshots its DB state under fine-grained locks, records the sequence-number bounds of the memtable to be flushed, and ensures the WAL is persisted up to that sequence before offloading the flush. {\color{black}Then, the owning LSM-KVS builds a compact flush-metadata package (572 bytes in total)}, which includes options (e.g., compression and block settings), file metadata (e.g., target level, file-number), KVS status metadata (e.g., column-family ID and sequence watermark), and WAL status (e.g., log file ID and persisted offset). The package also encodes DM locations for the index-block and KV-shard blocks. Since many metadata objects are shared in LSM-KVS, we precompute a small dependency graph and traverse it in a fixed order during packing to eliminate duplicates. Finally, the owning LSM-KVS sends a \texttt{PREPARE} message with the flush-metadata package to the scheduler.

\myline{Phase 2: \color{black}Assigning} Upon receiving \texttt{PREPARE} message, the scheduler pushes the flush job into a FIFO queue and dispatches jobs in arrival order to ensure fairness. For each job, it consults its active executor pool and chooses an execution mode: \emph{local} (the owning LSM-KVS executes the flush), \emph{in-DM} (a DM node executes in place without reading back to a CN), or \emph{remote-CN} (a different CN executes the flush job by reading the memtable from DM). After selecting a target node, the scheduler sends an \texttt{ASSIGN} message to the selected executor and starts a timeout. The executor responds with \texttt{ACCEPT} or the scheduler reassigns the job to another eligible executor at a different node when the timeout happens. Upon \texttt{ACCEPT}, the scheduler transmits the flush-metadata package and waits for \texttt{COMMIT} or \texttt{ABORT} from the executor.

\myline{Phase 3: \color{black}Executing} After the executor receives the flush-metadata package, it fetches the referenced memtable data from DM using one-sided \texttt{RDMA\_READ}. The executor directly reads the KV-block without pointer reconstruction, and the minimal in-memory index is re-anchored via base-address relocation. The executor (flush thread) merges records in sequence-number order, builds the required index and filter blocks, applies compression if configured, and writes $L_0$ SST files to DS as specified by the flush-metadata. If a failure occurs before completion, the executor aborts the job and sends an \texttt{ABORT} message to the scheduler so it can be safely reassigned (no state has been published). When the flush outputs are durable in DS, the executor sends a \texttt{COMMIT} message to the scheduler with the produced file metadata for the owning LSM-KVS to update flush state.

\myline{Phase 4: \color{black}Finalizing} After receiving \texttt{COMMIT}, the scheduler forwards the produced file metadata to the owning LSM-KVS. 
{\color{black}The owning LSM-KVS validates the metadata, installs the new DB state into the Manifest, and removes the memtable from its metadata under the DB mutex before notifying the DM to reclaim space, ensuring in-flight reads either find the data or fall back to the DS reads.}
The owning LSM-KVS then sends \texttt{ACK} to the scheduler. The scheduler marks the job complete, removes it from the queue, and releases its tracking state and timers. This phase publishes results safely and frees resources promptly.

\myline{Flush Scheduler} 
\textcolor{black}{
The scheduler maintains a dynamic pool of flush executors across CNs and DM nodes to balance the flush loads and mitigate write bursts happening in certain CNs. The scheduler uses heartbeats for node registration and liveness monitoring. To balance the offloaded flush tasks, we employ a deterministic cost model where each node $i$ is assigned a load factor $\text{load}_i = w_{\text{cpu}} u^{\text{cpu}}_i + w_{\text{io}} u^{\text{io}}_i + w_{\text{queue}} u^{\text{queue}}_i$. Here, the weights ($w$) are pre-configured scaling factors that prioritize resources based on the hardware bottleneck (e.g., higher $w_{\text{io}}$ for I/O-bound environments). The utilization metrics ($u$) are normalized values: $u^{\text{cpu}}_i$ is the ratio of active flush threads to the total thread pool size, $u^{\text{io}}_i$ represents the current storage write throughput relative to the node’s peak calibrated bandwidth, and $u^{\text{queue}}_i$ measures the current number of pending shards against the maximum allowable queue depth.}

\textcolor{black}{
Based on these signals, the scheduler calculates the estimated cost of assigning a flush job of size $\textit{bytes}$ to node $i$ using the formula $\text{cost}_i = \textit{bytes} \cdot (1 + \text{load}_i)$. We apply a greedy policy to select the node with the lowest cost to execute the flush job. For example, if Node A has a low load of 0.1 and Node B is heavily congested at 0.9, a 64 MB shard would result in costs of 70.4 and 121.6, respectively, leading the scheduler to select Node A. To ensure robustness, the scheduler recomputes $\text{load}_i$ from fresh telemetry signals upon the completion of each shard, preventing transient mispredictions from accumulating. Multiple scheduler instances can be deployed for scalability, while more complex scheduling strategies are left for future work.
}

\subsection{Shard-Level Optimizations} \label{sec:shard} 
Sending a 64 MB block in a single \texttt{RDMA\_WRITE} can cause bursty bandwidth contention, especially alongside other RDMA traffic \cite{liu2021consistent}. Transferring such large memtables also limits parallelism and does not resolve the slow $L_0$ compaction problem. To address these, we propose to shard the memtable. Each memtable is partitioned into multiple non-overlapping key ranges (called \emph{shards}). The single KV-block is thus divided into multiple contiguous KV-shard blocks. At the same time, we propose to flush multiple shard blocks in the same key-range from different memtables together, which actually combines $L_0$ compaction with flush. This can improve the flush parallelism and address the $L_0$ compaction penalty.

\myline{Memtable Sharding} 
We partition the memtable key space into $2^k$ non-overlapping key ranges (shards) using the first $k$ bits of each key. Within each shard, KV pairs are appended into a contiguous region, forming a \emph{KV-shard block}. Each index-block node stores a k-bit \texttt{shard\_id} alongside the KV pointer, and KV pairs are addressed by their offsets within the KV-shard block. Once a shard block becomes immutable, it is transferred to DM independently, so we do not wait for the entire memtable to seal, which shortens wait time and increases concurrency. With this design, smaller KV-shard blocks can be transferred asynchronously and in parallel without pointer reconstruction, fully utilizing RDMA bandwidth. During pointer correction on DM, we use the starting address of the KV-shard block rather than the whole KV-block for base-address relocation.

\myline{Shard-Level Flush Offloading}
To further improve scalability and parallelism, as well as $L_0$ compaction efficiency, O$^3$-LSM uses shard-level flush. O$^3$-LSM materializes shard-level flush metadata (e.g., shard ID and KV-shard blocks offsets on DM) and redesigns flush to aggregate multiple KV-shard blocks from the same key range of different memtables at DM into a single flush job. This shifts from coarse memtable-level flushing to fine-grained shard-level operations, enabling flush and $L_0$ compaction to be executed together. Unlike traditional designs that trigger a flush when accumulated memtables reach a size limit, the shard-level flush is triggered only when the total size of its KV-shard blocks across immutable memtables exceeds one memtable size limit (e.g., 64 MB). This keeps the generated $L_0$ SST files approximately the same size as the memtables, with minimal key-range overlap, thereby mitigating the $L_0$ compaction penalty. The shard-level optimization also handles workload skew. The scheduler provides fine-grained control by assigning hot shards to the most suitable nodes, balancing CPU and network load across the cluster.

\begin{figure}
  \centering
  \includegraphics[width=1\linewidth]{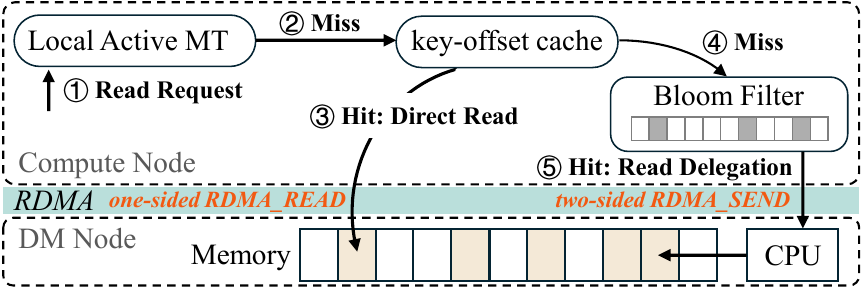}
  \caption{The workflow of \textit{Cache-Enhanced Read Delegation}.}
  \label{fig:figure5}
\end{figure}

\subsection{Cache-Enhanced Read Delegation} \label{sec:read delegation} 
In the original LSM-KVS design, responding to the read queries (e.g., \textit{Get} and \textit{Scan}) may involve traversing one or multiple memtables. Performing multiple one-sided \texttt{RDMA\_READ}s to traverse the memtable at DM involves: 1) dereferencing to obtain the key, 2) dereferencing to obtain the next node address based on key comparison results, and 3) repeating this process iteratively until the KV-pair is found (or NotFound), which can slow down read performance by an order of magnitude or more \cite{yoo2022read}. 

To address this multi-round traversal problem, two main optimization approaches can be considered: 1) reading data {\color{black}back} from DM to CN and conducting searching locally on the CNs (faster local processing, but searching the remote index-block can be slow), or 2) offloading searching tasks to the remote DM (asynchronized DM local search, but the computing resource might be limited). We combine those two approaches and propose \textbf{Cache-Enhanced Read Delegation} across local DRAM and remote DM. As shown in \autoref{fig:figure5}, for frequently accessed KV-pairs, we implement a local key-offset cache in O$^3$-LSM at CN, reducing the round-trip needed for memtable lookups. For less frequently accessed or newly accessed KV-pairs, we leverage the read delegation approach to utilize DM's computational capabilities for DM-local lookup, thereby reducing the multiple RDMA round-trip for searching the index-block. This method effectively combines the benefits of both one-sided and two-sided RDMA operations within a multi-tier architecture. 

\myline{Local key-offset cache} 
{\color{black}We designed a small, local key-offset cache for $O^3$-LSM. It uses an LRU-based policy to cache hot entries, mapping keys to their specific DM memory offsets.} We cache offsets rather than full KV-pair values to minimize CN memory usage, and the value will be fetched at cache hit with a single one-sided \texttt{RDMA\_READ}. The key-offset entries are inserted via the following two scenarios: 1) When an active memtable is cached in local DRAM, O$^3$-LSM tracks the read access frequency of each KV-pair, represented by \texttt{Freq} using 3 bits, which allows for 8 different frequency levels. When iterating the memtable during the transfer, key-offset entries are created for hot KV-pairs that meet a predefined frequency threshold (e.g., 4 or higher) and inserted into the cache. This approach enables one-sided \texttt{RDMA\_READ} to fetch the target KV-pair using the cached DM offset directly without searching the index-block. 2) When there is a lookup miss in the local key-offset cache, the KV-pair will be retrieved from DM via read delegation. The corresponding key-offset entry will be created and cached to expedite future requests for the same KV-pair. Additionally, O$^3$-LSM creates and caches a Bloom filter for each KV-shard block of the memtables stored in DM. This filter is used to validate whether a read delegation is needed or not, effectively reducing the average number of remote accesses.

\myline{Read delegation at DM} If a miss occurs in the local key-offset cache, we leverage DM's computational resources to delegate the search task. O$^3$-LSM first traverses the Bloom filters of the corresponding KV-shard blocks based on the \texttt{lookup\_key}'s \texttt{shard\_id}. If any Bloom filter returns positive, O$^3$-LSM packages the \texttt{lookup\_key} along with the set of memtable IDs into a read delegation request. Then, O$^3$-LSM executes an RPC call to DM using two-sided \texttt{RDMA\_SEND}. On the DM side, we maintain a worker pool with multiple polling threads that receive and process these read delegation requests. Upon receiving a request from CNs, a worker thread polls and performs the local lookup on the targeted memtables based on the request (delegated \textit{Get} operations) and transmits the result until the KV-pair is found (or NotFound if absent in all memtables). Once O$^3$-LSM receives the results from DM, we cache the key and its remote offset as a key-offset entry in the local key-offset cache.

\myline{Other optimizations} We proposed several RDMA optimizations: 1) \textbf{Doorbell batching.} We post 8–16 one-sided \texttt{RDMA\_READ}s at a time and ring the NIC doorbell once, which amortizes setup costs. 2) \textbf{Adaptive inlining for delegation replies.} For delegated lookups, the DM node returns small values directly within the RDMA reply. For larger values, it returns the metadata to perform a single follow-up one-sided read, effectively minimizing round-trip.

\begin{figure*}[!t]
  \begin{minipage}[b]{0.496\textwidth}
    \centering
    \includegraphics[width=\textwidth]{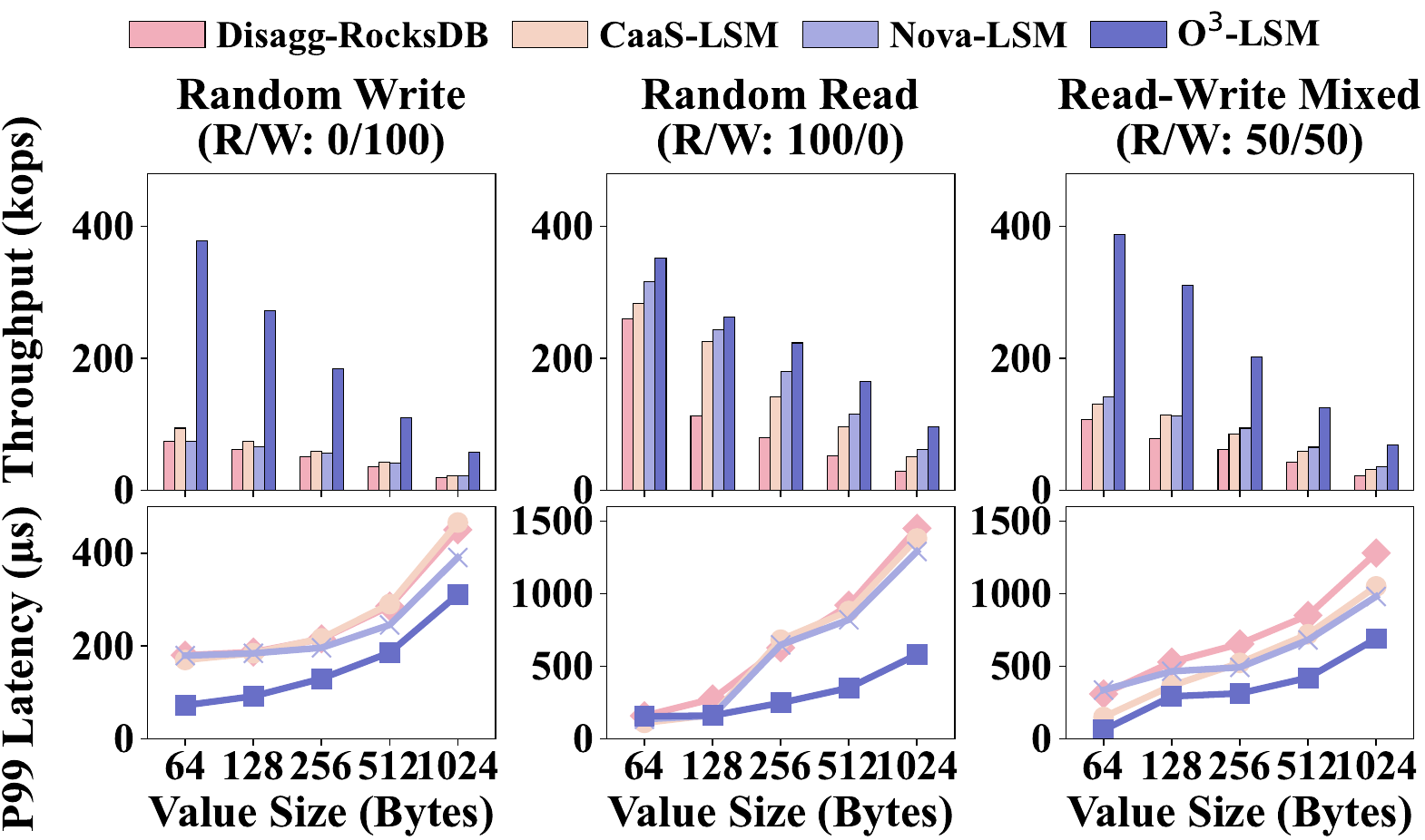}
    \subcaption{Performance under a 1 Gbps DS connection}
    \label{fig:1GB}
  \end{minipage}
  \hfill
  \begin{minipage}[b]{0.496\textwidth}
    \centering
    \includegraphics[width=\textwidth]{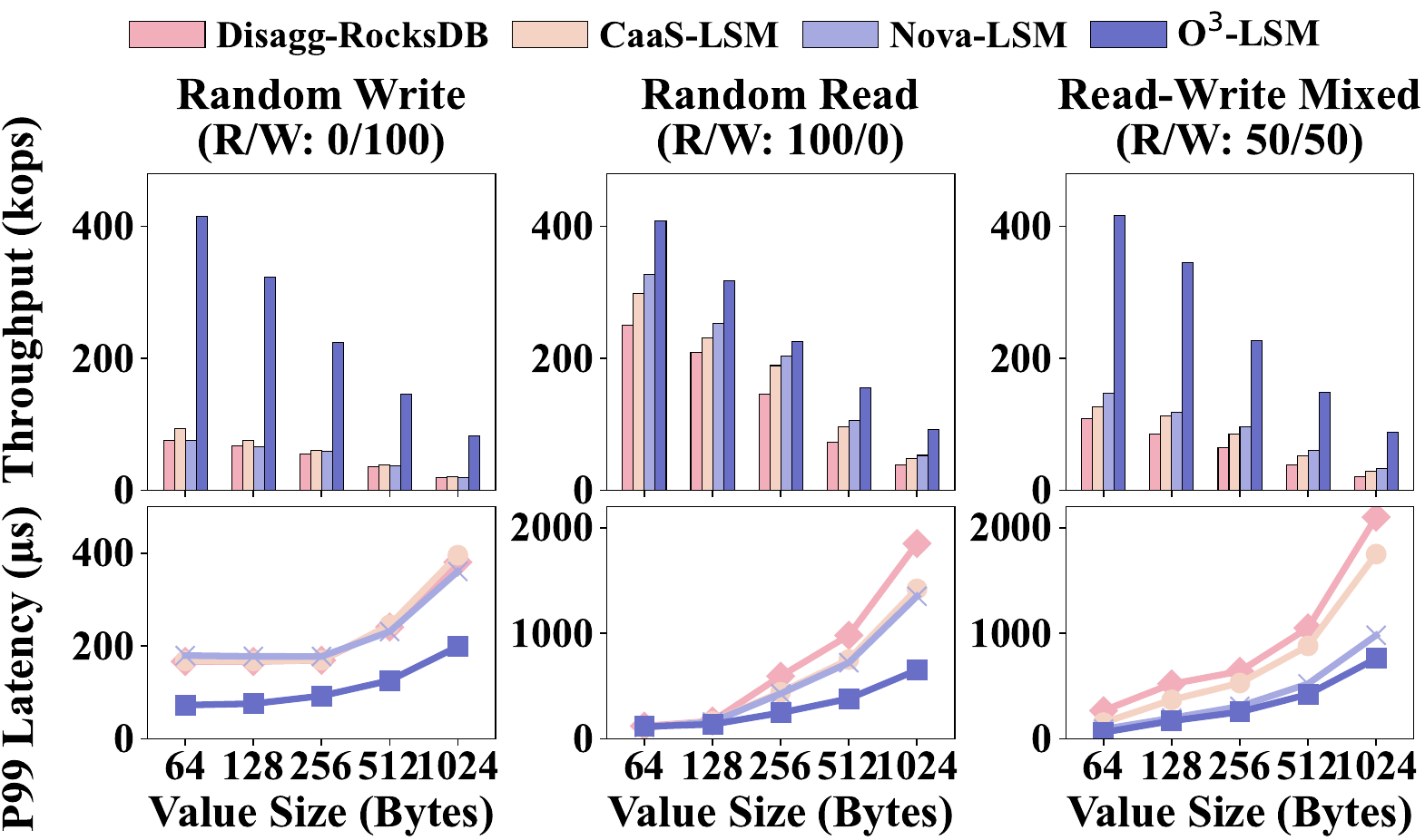}
    \subcaption{Performance under a 2.5 Gbps DS connection}
    \label{fig:2.5GB}
  \end{minipage}
  \caption{\color{black}Microbenchmark performance comparison under 1 Gbps and 2.5 Gbps DS connections.}
  \label{fig:microbenchmark}
  \vspace{-5pt}
\end{figure*}


\subsection{Fault Tolerance} 

\textcolor{black}{$O^3$-LSM ensures consistency by treating the CN’s WAL as the authoritative state and the Manifest as the sole visibility boundary.}

\noindent \textcolor{black}{
\myline{Job-Level Failures}
These are transient errors affecting individual operations:
1) Memtable Offloading: If an offload to DM fails, the CN releases the allocated DM region and rebuilds the memtable locally based on the WAL. No incomplete remote memtable is ever registered.
2) Shard-Level Flush: If a task fails during Flush execution, the scheduler discards partial SSTables and reschedules the shard using the immutable memtable on DM. Data only becomes visible after a successful Manifest update.
3) Read Delegation Failures: If a delegated read to DM fails or times out, the CN retries the request or, if the data has already been persisted, reads directly from the DS layer. This ensures that transient DM or network issues do not result in request failure.}

\noindent \textcolor{black}{
\myline{Node-Level Failures}
These involve the crash of entire system components:
1) CN Failure: On restart, the CN replays its local WAL to reconstruct memtables and re-offloads them to DM. The scheduler purges stale DM state associated with the failed CN.
2) DM Failure: The scheduler aborts all active jobs involving the failed DM. CNs rebuild the offloaded memtables from WAL locally.
3) Scheduler Failure: The scheduler recovers its state from a persistent log. It either resumes a recorded COMMIT or discards incomplete jobs, forcing CNs to retry.}
\textcolor{black}{
$O^3$-LSM maintains an all-or-nothing guarantee at the Manifest boundary. Partial outputs from failed jobs or nodes may necessitate retries, but they never result in partially installed SSTables or inconsistent metadata states.}

\section{\color{black}Evaluation} 
We implement O$^3$-LSM based on RocksDB v8.2.0 and Disaggregated RocksDB. The source code is available on GitHub \cite{O3-LSM}. We conduct comprehensive evaluations to answer the following questions: 1) Does O$^3$-LSM achieve explicit higher performance than state-of-the-art disaggregated LSM-KVS?  2) What is the performance breakdown of each major design? And 3) How about the performance with real-world applications, and scalability?

\subsection{Experimental Setup}
\myline{Hardware Platform} We conduct our evaluation on CloudLab's \cite{cloudlab} \texttt{c6220} instances, each equipped with 2 × Xeon E5-2650v2 CPUs (8 cores each), 64 GB of memory, and a 1 TB hard disk. To emulate the disaggregated setup, CNs are configured to use all available cores but only 2 GB of memory, whereas {\color{black}the DM node} are configured to use 4 cores while utilizing 64 GB of memory. All CNs and {\color{black} the DM node} are connected via Mellanox FDR ConnectX-3 NICs (40 Gbps connections). We deploy HDFS on the storage nodes as TCP-based DS (the same DS configuration as that in CaaS-LSM).

\myline{Baselines} To demonstrate the effectiveness of our designs, we use RocksDB optimized for DS \cite{dong2023rocksdb} as our baseline (\textbf{Disagg-RocksDB}). We also include \textbf{CaaS-LSM} \cite{yu2024caas-lsm} (a state-of-the-art LSM with remote compaction on DS) and \textbf{Nova-LSM} \cite{Huang_2021} (a disaggregated LSM-KVS with RDMA for DS accesses). All three baselines configure up to 8 local memtables at CN (totaling 512 MB).  In contrast, our design, \textbf{O$^3$-LSM}, uses only 2 local memtables (128 MB) and up to 6 memtables at the remote DM node. In addition, we also evaluate \textbf{dLSM} \cite{wang2023dlsm} (a fully in-memory LSM-KVS on RDMA-based DM) and use it as a performance upper-bound baseline (i.e., it is non-persistent and expected to achieve the best performance with RDMA-based DM).

\myline{Workloads} We use \texttt{db\_bench} and \textbf{YCSB} \cite{ycsb} to evaluate O$^3$-LSM. For \texttt{db\_bench}, we use \textit{fillrandom}, \textit{readrandom}, and \textit{readrandomwriterandom} benchmarks. For YCSB, we use workloads with 4 types of read-write ratios and 2 types of key-value distributions (\texttt{Uniform} and \texttt{Zipfian}). In Zipfian, keys are selected according to a Zipfian distribution, with skewness set to 0.99. Unless otherwise stated, all experiments are conducted with 1 DM node, 1 CN, and 3 DS nodes. Scalability and breakdown evaluations will use more CNs. We use the same configuration for O$^3$-LSM and other baselines. For memory components, the Memtable size is set to 64 MB, the block cache size is 512 MB, and the local Key-offset cache size is 64 MB. \textit{max\_background\_jobs} for compaction and flush is set to 4. The \textit{slowdown\_writes\_trigger} for $L_0$ is set to 32, and the \textit{stop\_writes\_trigger} is set to 48. {\color{black}The scheduling weights $w_{\text{cpu}}$, $w_{\text{io}}$, and $w_{\text{queue}}$ are all set to 1.0.} For NovaLSM, we configure \textit{num\_memtable\_partitions} to 12 to maximize parallelism and fully utilize RDMA bandwidth. We perform 50 million operations for each benchmark, with key/value sizes set to 16 bytes and 64 bytes, respectively. All evaluations are executed 3 times, and we present the average number.


\subsection{Overall Performance Evaluation}
We use \texttt{db\_bench} micro-benchmark to compare the overall performance of O$^3$-LSM with three baselines under limited network I/O (1 Gbps) and abundant I/O (2.5 Gbps) for accessing DS, as shown in \autoref{fig:1GB} and \autoref{fig:2.5GB} respectively. Also, we assess the performance of O$^3$-LSM with varying read-write ratios, key-value distributions via YCSB micro-benchmarks, and range queries.

\vspace{-4pt}
\myline{Write Performance} We first focus on intensive random write benchmarking with varying value sizes. We use Operations Per Second (Ops/s) to measure the throughput. As shown in \autoref{fig:1GB}, under 1 Gbps network bandwidth, O$^3$-LSM achieves up to 4.1$X$ throughput improvement over \textit{Disagg-RocksDB}, 3.2$X$ improvement over \textit{CaaS-LSM}, and 3.9$X$ improvement over \textit{Nova-LSM} (up to 6.4$X$, 4.6$X$, and 5.6$X$, respectively, for using 96 memtables in total). The P99 latency is also reduced by approximately 32\% to 59\%, demonstrating the significant write performance improvement of O$^3$-LSM. \textit{Disagg-RocksDB} suffers throughput drops with larger value sizes due to increased write pressure, straining network I/Os to DS, and $L_0$ compaction. 
Similarly, O$^3$-LSM shows greater throughput improvement over \textit{CaaS-LSM} as value sizes grow. This is because as value sizes increase, O$^3$-LSM’s \textit{Shard-Level Flush Offloading} results in lower write amplification and finer-grained scheduling, enhancing overall performance. For \textit{Nova-LSM}, throughput drops as value sizes increase, while P99 latency escalates sharply. Despite using RDMA to interconnect compute and storage clusters, the overhead of managing larger data blocks across the network eventually outweighs the bandwidth benefits provided by its sharded architecture. Notably, O$^3$-LSM outperforms \textit{Nova-LSM} in both throughput and P99 latency with much slower TCP-based DS.

\autoref{fig:2.5GB} shows the random write performance evaluation under abundant bandwidth (2.5 Gbps DS connections). All four disaggregated LSM-KVS achieve better performance compared to the limited 1 Gbps network bandwidth. As the value size increases, O$^3$-LSM achieves up to 4.5$X$, 3.4$X$, and 4.4$X$ throughput improvements over \textit{Disagg-RocksDB}, \textit{CaaS-LSM}, and \textit{Nova-LSM}, respectively (up to 6.9$X$, 5.7$X$, 6.2$X$ for using 96 memtables in total). Additionally, by increasing the CN to DS network bandwidth from 1 Gbps to 2.5 Gbps, the P99 latency of O$^3$-LSM decreases by approximately 37\% to 60\%, indicating that O$^3$-LSM can deliver better performance under abundant bandwidth.
\textcolor{black} {As a high-throughput, non-persistent LSM reference, dLSM reaches 1.23 million ops (Mops) at 64B value size and is omitted from the figure for clarity. As value sizes increase from 64B to 1024B, OPS across all baselines drops because the fixed network bandwidth creates a direct trade-off with payload size; similarly, dLSM’s throughput scales from 1.23 to 0.58 Mops. Against this reference at 64B, O$^3$-LSM attains 0.415 Mops (34\% of dLSM), whereas other persistent LSM-KVS baselines only achieve up to 0.09 Mops (7\% of dLSM).}

\myline{Read Performance} We evaluated the point lookup performance using the \textit{readrandom} benchmark from \texttt{db\_bench}. First, we loaded 50 million records and waited for all compactions to complete before conducting the \textit{readrandom} test. As shown in \autoref{fig:1GB} and \autoref{fig:2.5GB}, all systems achieved slightly better read throughput under abundant bandwidth compared to limited bandwidth. For P99 latency, both \textit{CaaS-LSM} and \textit{Nova-LSM} showed significant reductions of 35\% and 33\%, respectively, under abundant bandwidth with large value sizes. O$^3$-LSM reduced P99 latency by up to 69\% compared to the other baselines under limited bandwidth. O$^3$-LSM demonstrated up to 1.8$X$, 0.6$X$, and 0.3$X$ read throughput improvements over \textit{Disagg-RocksDB}, \textit{CaaS-LSM}, and \textit{Nova-LSM}, respectively (up to 2.2$X$, 1.8$X$, 1.5$X$ for using 96 memtables in total). However, as the value size increases, the performance of all systems declines, particularly for O$^3$-LSM. This is because, although Cache-Enhanced Read Delegation accelerates read performance when accessing memtables stored on DM, larger value sizes consume more RDMA bandwidth, leading to a performance drop. Compared with dLSM, O$^3$-LSM achieves up to 11\% of the upper bound 
\textcolor{black} {(2.24 Mops at 64B and 1.25 Mops at 1024B) }.
Other baselines can achieve only 0.2 Mops (8.9\% of dLSM). In general, Cache-Enhanced Read Delegation effectively mitigates remote-memory access overhead for memtable hits at DM in O$^3$-LSM. 
Crucially, Shard-Level Flush Offloading produces far fewer $L_0$ SSTables with overlapping key ranges. Since most data resides on DS, the majority of read queries are handled there. Consequently, O$^3$-LSM transforms $L_0$ lookups from probing multiple overlapping files into fewer probes for point queries and reduces iterator fan-in for scans. This structural reduction in read amplification eliminates redundant Bloom or metadata checks while lowering CPU and I/O overhead. Therefore, this design achieves better read performance compared to baselines by minimizing DS reads.

\myline{Read-Write Mixed Workload}  We use the \textit{readrandomwriterandom} benchmark. We evaluate a 50:50 read–write ratio under varying bandwidth pressure. As shown in \autoref{fig:1GB} and \autoref{fig:2.5GB}, O$^3$-LSM achieves the best performance in both cases. Compared to \textit{Disagg-RocksDB}, \textit{CaaS-LSM}, and \textit{Nova-LSM}, O$^3$-LSM achieves up to 3$X$, 2.3$X$, and 1.9$X$ overall throughput improvements, respectively (up to 3.9$X$, 3.6$X$, 3$X$ for using 96 memtables in total), and reduces the P99 latency up to 76\%. O$^3$-LSM achieves up to 16.5\% of the dLSM throughput 
\textcolor{black} {(1.53 Mops at 64B and 0.82 Mops at 1024B) }
while other baselines only achieve lower than 10\% of dLSM.

\begin{figure}[t]
  \centering
  \begin{minipage}[t]{0.48\textwidth}
    \centering
    \includegraphics[width=\linewidth]{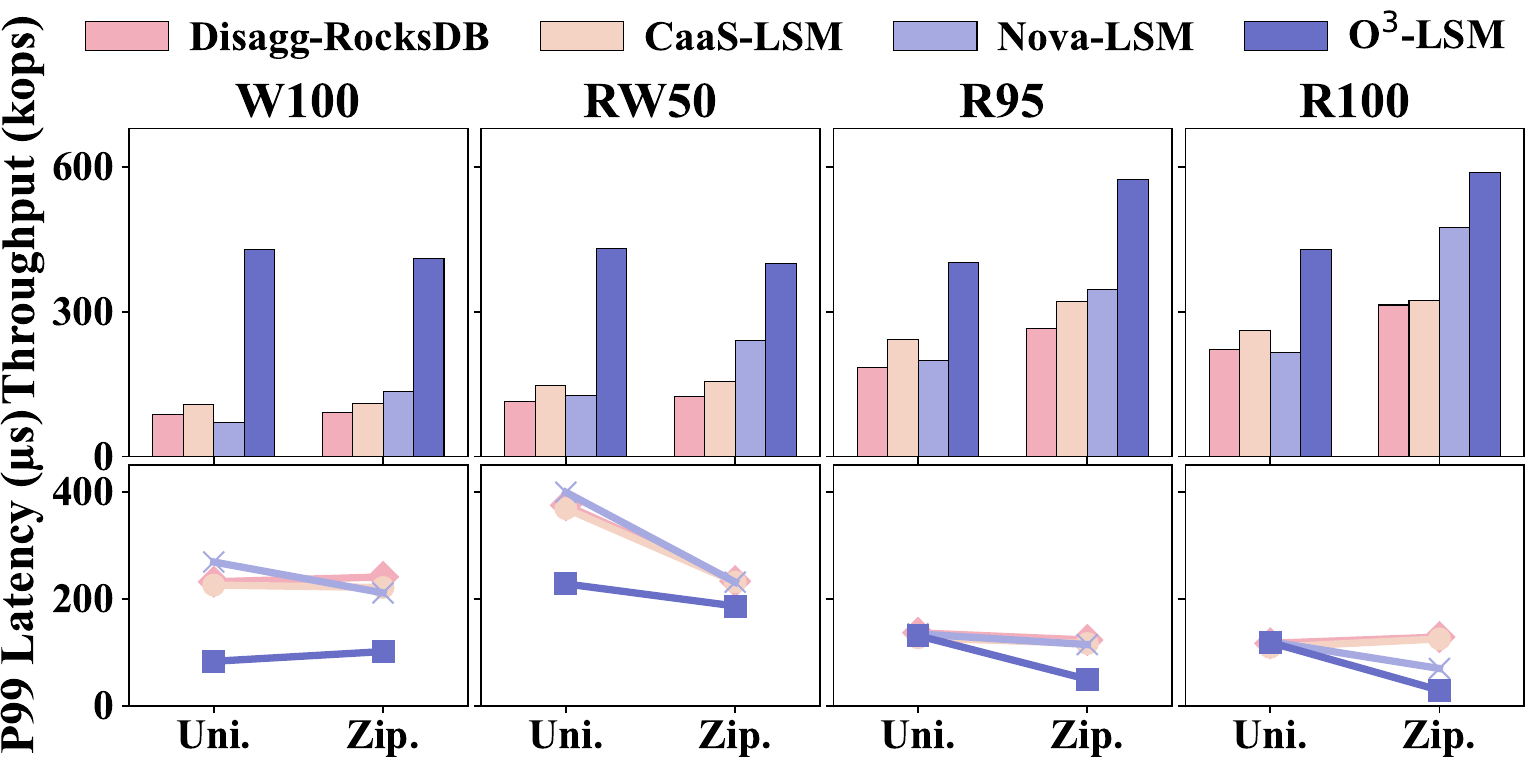}
    \captionsetup{labelformat=parens, labelsep=space}
  \end{minipage}
  \caption{YCSB {\color{black}micro}-benchmark performance.}
  \label{fig:yscb}
\end{figure}


\begin{figure}[t]
  \centering
  \begin{minipage}[t]{0.48\textwidth}
    \centering
    \includegraphics[width=\linewidth]{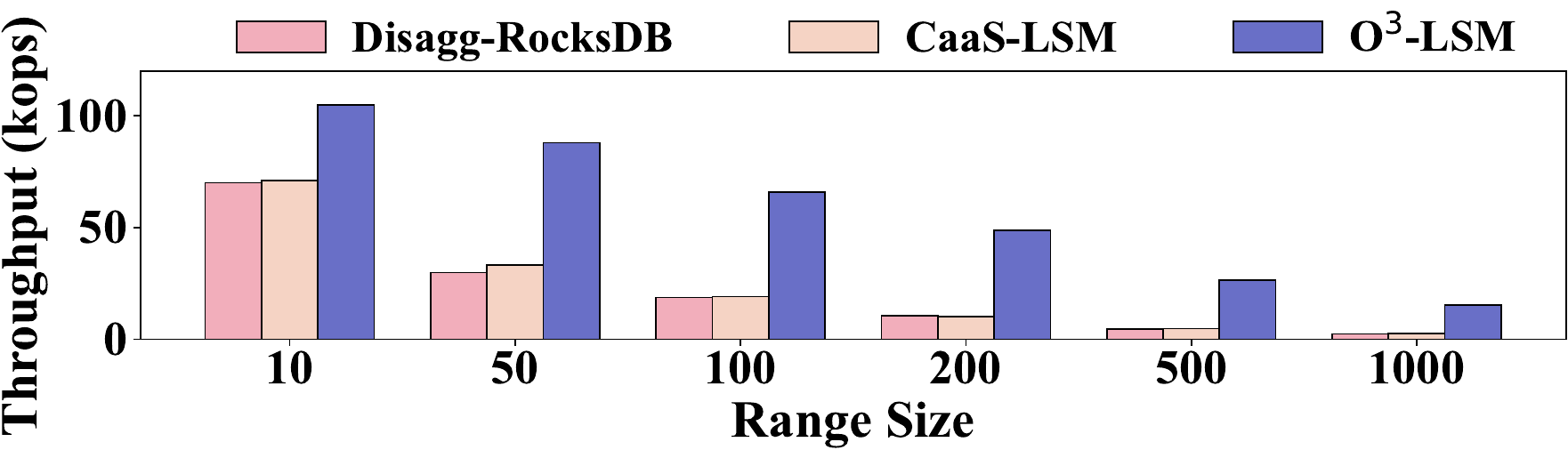}
    \captionsetup{labelformat=parens, labelsep=space}
  \end{minipage}
  \caption{Range query performance.}
  \label{fig:range}
  \vspace{-5pt}
\end{figure}

\myline{{\color{black}Micro}-benchmark} 
We used YCSB, a widely adopted benchmark, in this evaluation. We employed \textit{Uniform} and \textit{Zipfian} key distributions and evaluated 4 different read-write ratios. We focused on four workloads: 1) \textbf{RW50} (50\% read and 50\% write), 2) \textbf{R95} (95\% read and 5\% write), 3) \textbf{R100} (100\% read), 4) \textbf{W100} (100\% write) with a value size of 64 bytes and 2.5Gbps sufficient bandwidth. \autoref{fig:yscb} shows that O$^3$-LSM achieves the best performance across all evaluations. We draw two main conclusions: First, the optimizations in O$^3$-LSM are more effective under the \textit{Zipfian} distribution compared to the \textit{Uniform} distribution, particularly for read-intensive workloads (R95 and R100). \textit{Zipfian} generates hot KV-pairs, and the Key-offset Cache can effectively cache them in CN local memory, which significantly reduces the latency of accessing DS compared to the other three baselines. Second, O$^3$-LSM shows a slight performance drop in write-intensive workloads (W100) under the \textit{Zipfian} distribution. The skew concentrates most inserts into a small set of shards, which impacts the shard-level parallelism and causes a lot of key-range overlapped SST files in $L_0$.

\myline{Range Query} We evaluate the performance of range queries using different range sizes. As shown in \autoref{fig:range}, O$^3$-LSM outperforms both \textit{Disagg-RocksDB} and \textit{CaaS-LSM}. As the range size increases (e.g., from 10 to 1000), the throughput of all three systems decreases. Differently, O$^3$-LSM achieves significantly better throughput improvements with up to 5.2$X$ and 4.6$X$ compared to \textit{Disagg-RocksDB} and \textit{CaaS-LSM}, respectively. This improvement is due to two key factors: 1) O$^3$-LSM performs shard-level flush, which benefits range queries by assigning many consecutive keys to the same shard, significantly reducing prefetch time when reading SST files from DS. And 2) O$^3$-LSM saves the local memory for a larger block cache, which further reduces the DS I/Os.

\vspace{-2pt}
\subsection{Performance Breakdown Analysis} \label{sec:performance-break}
\begin{figure}
\centering
\begin{subfigure}{.485\columnwidth}
  \centering
  \includegraphics[width=\linewidth]{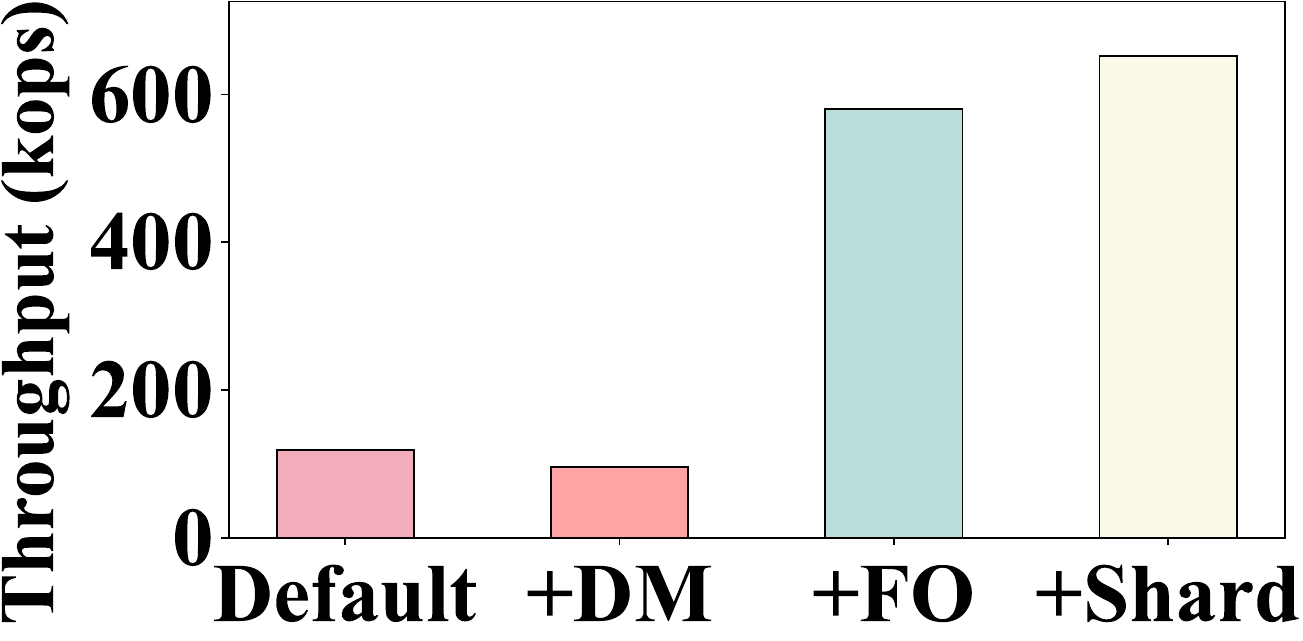}
  \caption{Throughput}
  \label{subfig:flush-breakdown}
\end{subfigure} \hfill
\begin{subfigure}{.485\columnwidth}
  \centering
  \includegraphics[width=\linewidth]{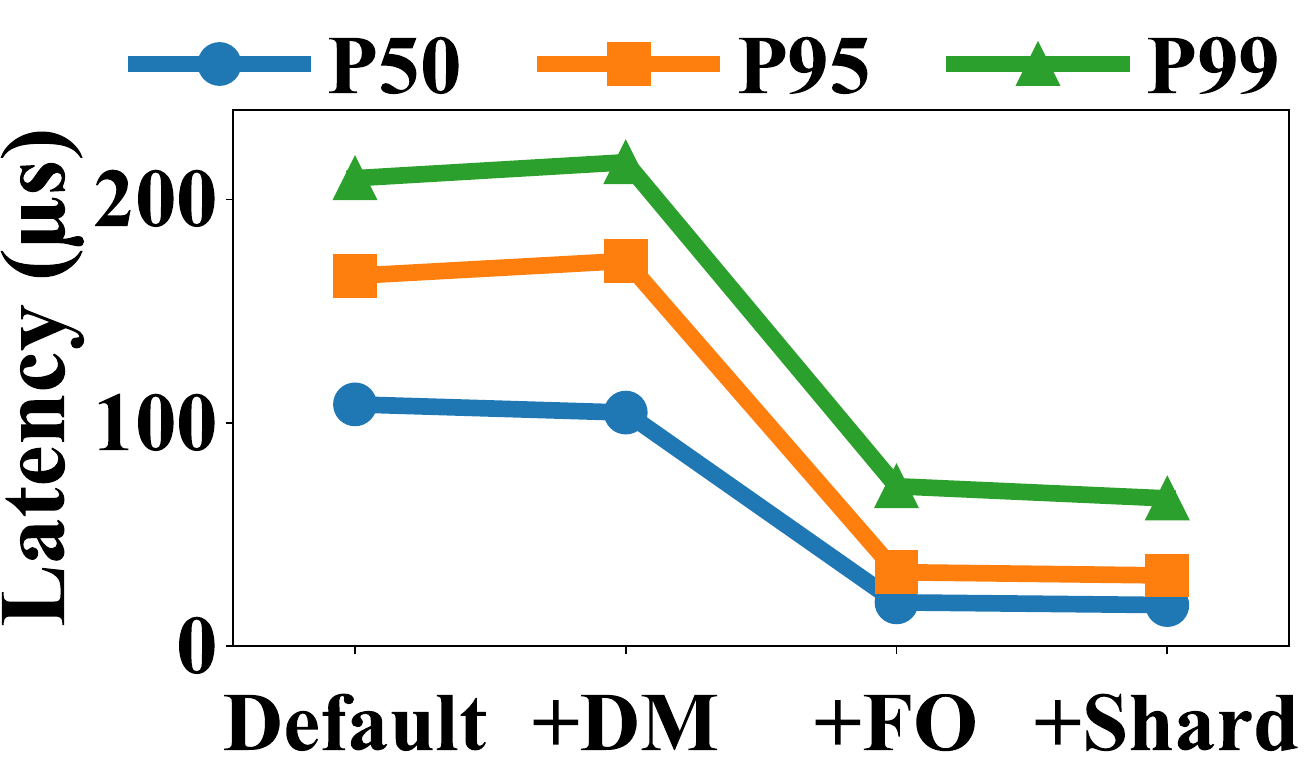}
  \caption{Latency}
  \label{subfig:flush-breakdown-latency}
\end{subfigure}
\caption{Ablation study on the \textit{fillrandom} workload.}
\label{fig:breakdown}
\end{figure}

\myline{Write Performance} As shown in \autoref{subfig:flush-breakdown}, simply enabling disaggregated memory (\textbf{+DM}) to extend the write buffer does not alleviate the write slowdown problem caused by the limited write buffer. In fact, it has about 18\% throughput decrease and slight latency increase compared with \textit{Disagg-RocksDB} (\textbf{Default}), due to its slow memtable transfer/rebuild and explicit flush operation delay. The performance can be even worse when flushing more memtables at DM concurrently. When the number of memtables to be flushed exceeds 10, flushing the memtables at DM is slower than the original \textit{Disagg-RocksDB}. However, when the proposed \textit{Flush Offloading} (\textbf{+FO}) is integrated, the flush throughput increases 34.6$X$, improved from 5MB/s to 178MB/s (100 memtable limit). More importantly, collaborative flush offloading can achieve high scalability (i.e., as the number of memtables increases, the aggregated flush throughput also increases almost linearly). Furthermore, by breaking down \textit{Flush Offloading} into shard-level operations (\textbf{+Shard}), we achieve even better overall write throughput (652K Ops/s) and higher flush throughput (15\% improvement compared with \textbf{+FO}).

\begin{table}[t]
\centering
\resizebox{\columnwidth}{!}{
\begin{tabular}{@{}llcccc@{}}
\toprule
\textbf{Workload} & \textbf{System} & \textbf{L0 Comp.} & \textbf{L1 Comp.} & \textbf{L2 Comp.} & \textbf{Total} \\ \midrule
Random Write & $O^3$-LSM & 12.6 & 18.4 & 19.1 & 50.1 \\
(50M, 256B)  & Disagg-RocksDB   & 12.7 & 22.4 & 24.2 & 59.3 \\ \midrule
Mixed (50/50)       & $O^3$-LSM & 6.3  & 9.2  & 16.4 & 31.9 \\
(50M, 256B)  & Disagg-RocksDB  & 6.4  & 12.1 & 21.3 & 39.8 \\ \bottomrule
\end{tabular}
}
\caption{\color{black}Write Amplification Analysis (Data written in GB).}
\label{tab:wa-analysis}
\vspace{-5pt}
\end{table}

\noindent \textcolor{black}{
\textbf{Write Amplification Analysis.} \autoref{tab:wa-analysis} provides a breakdown of data written during compaction across different LSM levels. $O^3$-LSM consistently achieves lower total write amplification (WA) compared to the Disagg-RocksDB. Specifically, for the 50M random write workload, $O^3$-LSM reduces the total data written by approximately 15.5\%. This reduction is primarily observed in $L_1$ and $L_2$ compaction. By utilizing shard-level flush offloading and memtable offloading, $O^3$-LSM generates fewer overlapping $L_0$ SST files, which significantly reduces the overlapping key ranges during subsequent compaction levels, thereby mitigating the overall WA.}


\myline{Flush Performance Breakdown}
As shown in \autoref{tab:flush_comparison_final}, compared with directly flushing the memtable from local memory to DS (Local Baseline), O$^3$-LSM’s \textit{Shard-Level Flush Offloading} adds only 7\% average end-to-end latency. In the overall timeline, transferring a memtable to DM contributes 18.7\% of total time, but our asynchronous memtable transfer design decouples this cost from the flush critical path by overlapping transfer with execution. Within the \textit{Flush Offloading} critical path, flush metadata packaging/parsing (\textbf{MP}) averages 4.7ms, install/finalization (\textbf{IF}) averages 3.2ms, and the executor’s fetch of metadata and memtable blocks (\textbf{FM}) accounts for 5.7\% of total execution time (54ms), the remainder is spent on merge/serialize/write to DS (\textbf{Flush}). These results indicate that the proposed collaborative flush offloading protocol overheads are modest and largely hidden by overlap, while the dominant costs lie in data movement and DS I/O.

\begin{table}[t]
\centering
\small
\renewcommand{\arraystretch}{1.1}
\setlength{\tabcolsep}{4pt}
\begin{tabular}{@{}l rrrr r@{}}
\toprule
\textbf{Method} & \textbf{MP} & \textbf{Flush} & \textbf{IF} & \textbf{FM/Other} & \textbf{Total (ms)} \\
\midrule
Disagg-RocksDB & 3.2 & 877 & 2.5 & N/A & 882.7 \\
CaaS-LSM       & 3.1 & 874 & 2.6 & N/A & 879.7 \\
Nova-LSM       & 1.9 & 852 & 2.1 & N/A & 856.0 \\
\midrule
Naive DM Solution & 3.2 & 877 & 2.5 & 286.0$^{\dagger}$ & 1168.7 \\
\textbf{O$^3$-LSM} & \textbf{4.7} & \textbf{875} & \textbf{3.2} & \textbf{54.0} & \textbf{936.9} \\
\bottomrule
\multicolumn{6}{l}{\scriptsize $^{\dagger}$ Includes Transfer and rebuild (218 ms), and Retrieval (68 ms) from the naive implementation.}
\end{tabular}
\caption{\color{black}Flush latency breakdown analysis.}
\label{tab:flush_comparison_final}
\end{table}

\myline{Read Performance}
As shown in \autoref{tab:read}, we compare O$^3$-LSM with several baselines to highlight the effectiveness of its read performance optimizations. First, we introduce a straightforward approach that O$^3$-LSM uses multiple one-sided \texttt{RDMA\_READ} operations to search memtables at DM (\textbf{One-side Read}). One-sided read can only achieve 8,426 Ops/s and cause extremely high latency (547 us), which is even significantly worse than \textit{Disagg-RocksDB}. It shows that directly using one-sided \texttt{RDMA\_READ} operations to search memtables at DM can lead to extremely high overhead. Second, when we enable \textit{Read Delegation} (\textbf{RD}) in O$^3$-LSM, it outperforms \textit{Disagg-RocksDB}, delivering 0.23X and 0.27X higher throughput and 80\% and 85\% lower P50 latency under Uniform and Zipfian distributions, respectively. Finally, with the addition of a local key-offset cache (\textbf{RD+LC}) as the complete cache-enhanced read delegation design, compared with RD only scheme, \textbf{RD+LC} achieves a 15\% and 35\% improvement in throughput, along with a 2\% and 19\% reduction in latency, under Uniform and Zipfian distributions, respectively.

\begin{table}[t]
\centering

\setlength{\tabcolsep}{4pt}
\renewcommand{\arraystretch}{1.05}

\begin{tabular}{@{}lrrrr@{}}
\toprule
& \multicolumn{2}{c}{\textbf{P50 Latency (µs)}} & \multicolumn{2}{c}{\textbf{Throughput (KOps/s)}} \\
\cmidrule(lr){2-3}\cmidrule(lr){4-5}
\textbf{Baselines} & \textbf{Uniform} & \textbf{Zipfian} & \textbf{Uniform} & \textbf{Zipfian} \\
\midrule
One-side Read    & 547.369 & 501.327 &   8.426 &  11.265 \\
Disagg-RocksDB  &  37.465 &  31.129 & 296.975 & 325.315 \\
RD              &   5.425 &   6.941 & 366.975 & 414.791 \\
RD+LC           &   5.370 &   5.597 & 424.460 & 559.711 \\
\bottomrule
\end{tabular}
\caption{Read performance breakdown (Uniform \& Zipfian).}
\label{tab:read}
\end{table}

\begin{table}[t]
\centering
\resizebox{\columnwidth}{!}{
\begin{tabular}{@{}lcccc@{}}
\toprule
\textbf{Workload / System} & \textbf{CN$\rightarrow$DM} & \textbf{DM$\rightarrow$CN} & \textbf{CN$\rightarrow$DS} & \textbf{DM$\rightarrow$DS} \\ \midrule
\textit{Random Read ($O^3$-LSM)}  & 1.1 & 6.7 & 4.2 & 0.0 \\
\textit{Random Read (Disagg-RocksDB)} & N/A & N/A & 10.7 & N/A \\ \midrule
\textit{Random Write ($O^3$-LSM)} & 12.6 & 8.4 & 46.1$^{\ddagger}$ & 4.3 \\
\textit{Random Write (Disagg-RocksDB)}& N/A & N/A & 59.3 & N/A \\ \midrule
\textit{Mixed ($O^3$-LSM)}        & 6.3 & 5.2 & 29.6$^{\ddagger}$ & 2.4 \\
\textit{Mixed (Disagg-RocksDB)}      & N/A & N/A & 39.8 & N/A \\ \bottomrule
\multicolumn{5}{l}{\small $^{\ddagger}$ Traffic is split between two CNs: \textbf{Random Write} (25.36/20.74 GB), \textbf{Mixed} (19.5.2/10.1 GB).}

\end{tabular}
}
\caption{\color{black}Network I/O Breakdown (Data volume in GB).}
\label{tab:network-io}
\vspace{-5pt}
\end{table}

\begin{figure}
\centering
\begin{subfigure}{.485\columnwidth}
  \centering
    \includegraphics[width=\textwidth]{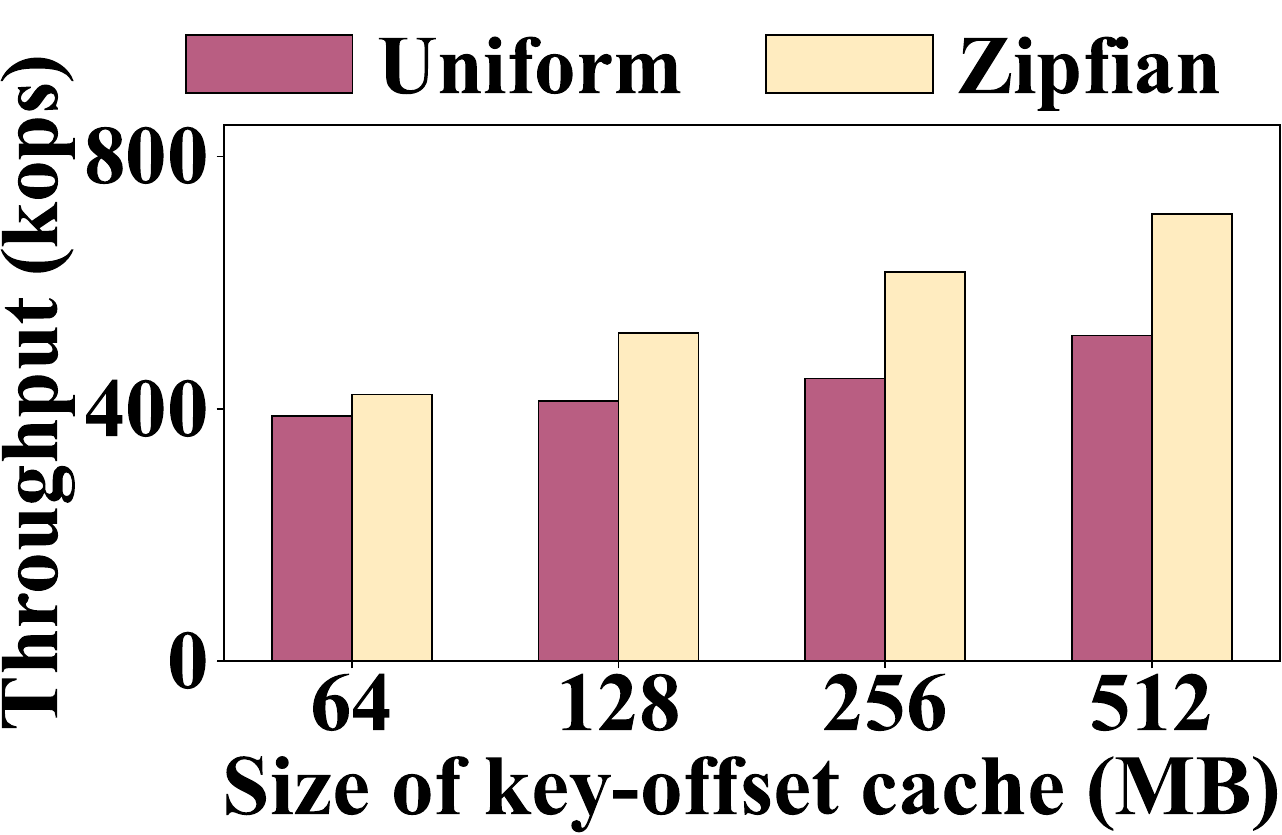}
    \subcaption{Varying key-offset cache}
    \label{fig:breakcachesize}
\end{subfigure} \hfill
\begin{subfigure}{.485\columnwidth}
  \centering
    \includegraphics[width=\textwidth]{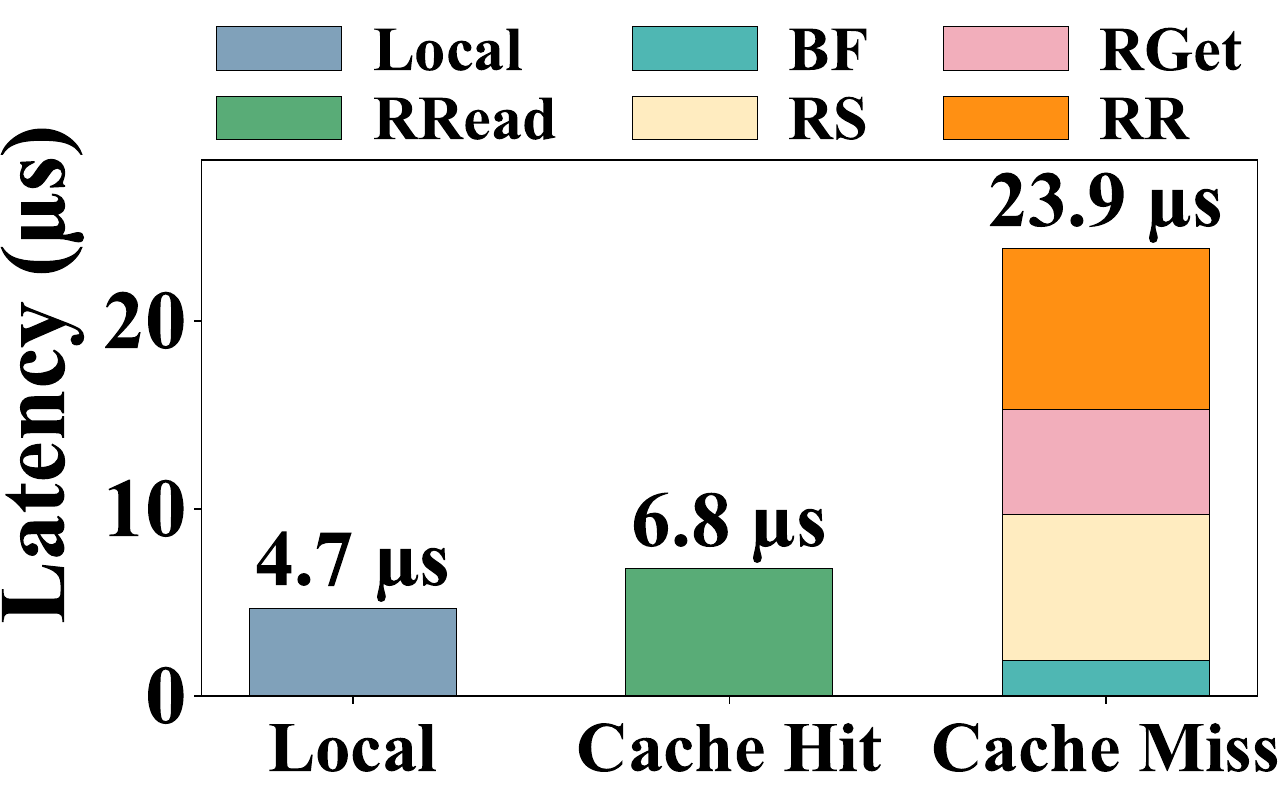}
    \subcaption{Time breakdown}
    \label{subfig:breakhit}
\end{subfigure}
  \caption{Read performance analysis.}
  \label{fig:breakdown-read}
\end{figure}

\myline{Key-Offset Cache Size} To analyze the impact of the key-offset cache size on the read performance, we conducted experiments using Uniform and Zipfian \textit{readrandom} workloads. As shown in \autoref{fig:breakcachesize}, the performance of O$^3$-LSM improves as the key-offset cache capacity increases. Specifically, for the Zipfian distribution workload, increasing the cache size from 64 MB to 512 MB resulted in a 67\% throughput improvement. In contrast, for the uniform distribution workload, throughput increased by 32\%. This shows that our design can significantly improve read performance with a larger key–offset cache and is more effective for skewed workloads.

\myline{Read Time Breakdown}
As shown in \autoref{subfig:breakhit}, we present three read cases. The first scheme looks up memtables at local memory (\textbf{Local}) and costs 4.7us. The second and third correspond to our Cache-Enhanced Read Delegation. On a cache hit (\textbf{Cache Hit}), a single one-sided \texttt{RDMA\_READ} completes in 6.8us (\textbf{RRead}). On a cache miss (\textbf{Cache Miss}), the pipeline performs a Bloom-filter check (\textbf{BF}, 1.89us), then \texttt{RDMA\_SEND} to the DM node (\textbf{RS}, 7.8us), followed by the remote search on DM (\textbf{RGet}, 5.6us, higher than local due to extra copies), and finally \texttt{RDMA\_RECV} of the result (\textbf{RR}, 8.6us), for a total of 23.9us. Both two-sided RDMA operations are amortized via doorbell batching, which lowers effective cost. Relative to the naive solution at 29us, the miss path reduces latency by 17.6\% (5.1us), and the cache-hit path reduces latency by 76.6\% (22.2us).


\begin{figure}[t]
  \centering
  \begin{minipage}[t]{0.48\textwidth}
    \centering
    \includegraphics[width=\linewidth]{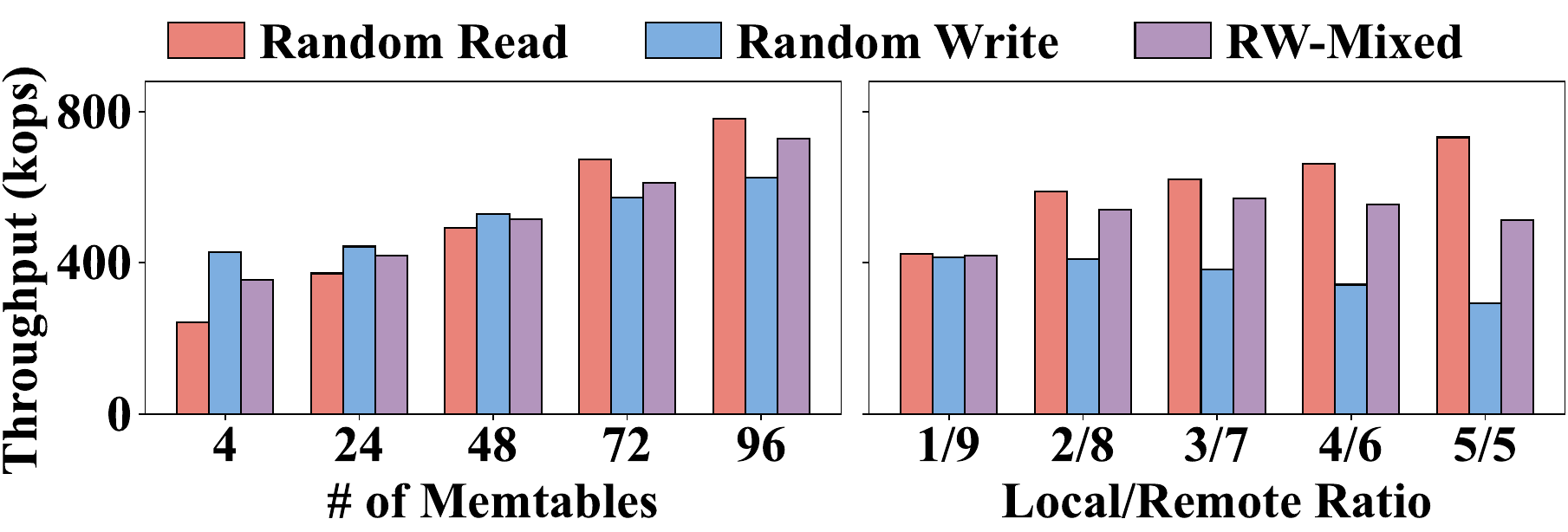}
    \captionsetup{labelformat=parens, labelsep=space}
  \end{minipage}
  \caption{Varying memtable count and local/remote ratio.}
  \label{fig:memtable}
  \vspace{-5pt}
\end{figure}

\vspace{-2pt}
\myline{Performance Influence of Different Maximum Memtables} \autoref{fig:memtable} shows the performance of various maximum numbers of memtables allowed in O$^3$-LSM (2 at most in CN and the rest will be transferred to DM). As more memtables can be stored in DM, O$^3$-LSM demonstrates improved performance across different workloads. This outcome aligns with our motivation: caching more memtables in the write buffer enhances overall performance. This result highlights the success of our design in overcoming the new challenges by extending the write buffer with DM. For a fixed maximum number of memtables in the write buffer (in our tests, this was set to 10 memtables), the ratio of local to remote memtables also impacts overall performance. As shown in \autoref{fig:memtable}, increasing the percentage of local memtables improves read performance since more KV-pairs can be searched in the local memtables. However, write performance declines since we are less efficient at leveraging other nodes (CNs or DM nodes) to flush these local memtables to the DS, resulting in decreased performance. Exploring the automatic memtable ratios between local and DM will be our future work.


\noindent  \textcolor{black}{
\textbf{Network I/O Breakdown.} As shown in \autoref{tab:network-io}, we analyze how $O^3$-LSM redistributes network traffic across different nodes. In \textit{Disagg-RocksDB}, the CN (owning LSM-KVS) handles all I/O for flushing data to DS. In contrast, $O^3$-LSM can offload a significant portion of this traffic to other available CNs or DM nodes. In our 2-CN evaluation of a 50M random write workload, although we introduce 12.6~GB of traffic for memtable offloading (CN$\rightarrow$DM), we successfully reduce the direct CN$\rightarrow$DS pressure from 59.3~GB to 46.1~GB. This traffic is effectively rebalanced across the two CNs and the DM node, with the two CNs contributing 25.36~GB and 20.74~GB respectively, while the DM node handles the remaining 4.3~GB. These results validate the effectiveness of our proposed scheduling algorithm in redistributing network I/O across components. This shift is critical during write-intensive bursts, as it prevents the network interface of a single node from becoming a bottleneck by distributing the network I/Os across all available resources.}

\begin{table}[t]
\centering
\resizebox{\columnwidth}{!}{
\begin{tabular}{@{}llcccc@{}}
\toprule
\textbf{System} & \textbf{Metric} & \textbf{CN$\rightarrow$DM} & \textbf{DM$\rightarrow$CN} & \textbf{CN$\rightarrow$DS} & \textbf{DM$\rightarrow$DS} \\ \midrule
$O^3$-LSM w/ Shard  & Peak    & 146.2  & 86.7   & 265.0 & 42.3 \\
                & Average & 58.4  & 34.2   & 234.0 & 40.2 \\ \midrule
$O^3$-LSM w/o Shard & Peak    & 963.2 & 216.5 & 297.0 & 259.0 \\
                & Average & 39.2  & 24.6   & 214.0 & 36.8 \\ \bottomrule
\end{tabular}
}
\caption{\color{black}Shard-Level Optimization on Network Traffic (MB/s).}
\label{tab:shard-impact}
\end{table}

\noindent \textcolor{black}{
\textbf{Shard-level optimization.} \autoref{tab:shard-impact} demonstrates the critical role of shard-level partitioning in smoothing network traffic. Without shard-level optimization, the system experiences massive traffic spikes, with peak CN$\rightarrow$DM throughput reaching 963.2~MB/s against an average of only 39.2~MB/s. This bursty behavior leads to severe network congestion. By partitioning memtables into shards and scheduling their transfer and flush in parallel, $O^3$-LSM reduces peak CN$\rightarrow$DM traffic by $6.5\times$ (from 963.2 to 146.2~MB/s), while increasing the average throughput. These results illustrate how sharded flush mitigates bandwidth bursts and smooths network traffic across nodes. {\color{black} Consequently, sharding boosts W100-Uniform throughput from 268.1 to 428.9 kops ($1.6\times$). Even under W100-Zipfian skew, $O^3$-LSM achieves 409.5 kops, which is a $1.7\times$ gain over the non-sharded baseline (241.2 kops). This demonstrates that parallelizing shard flushes mitigates hotspots and maintains high performance by allowing independent flush tasks to proceed concurrently.}
}

\begin{figure}[t]
  \centering
  \begin{minipage}[t]{0.45\textwidth}
    \centering
    \includegraphics[width=\linewidth]{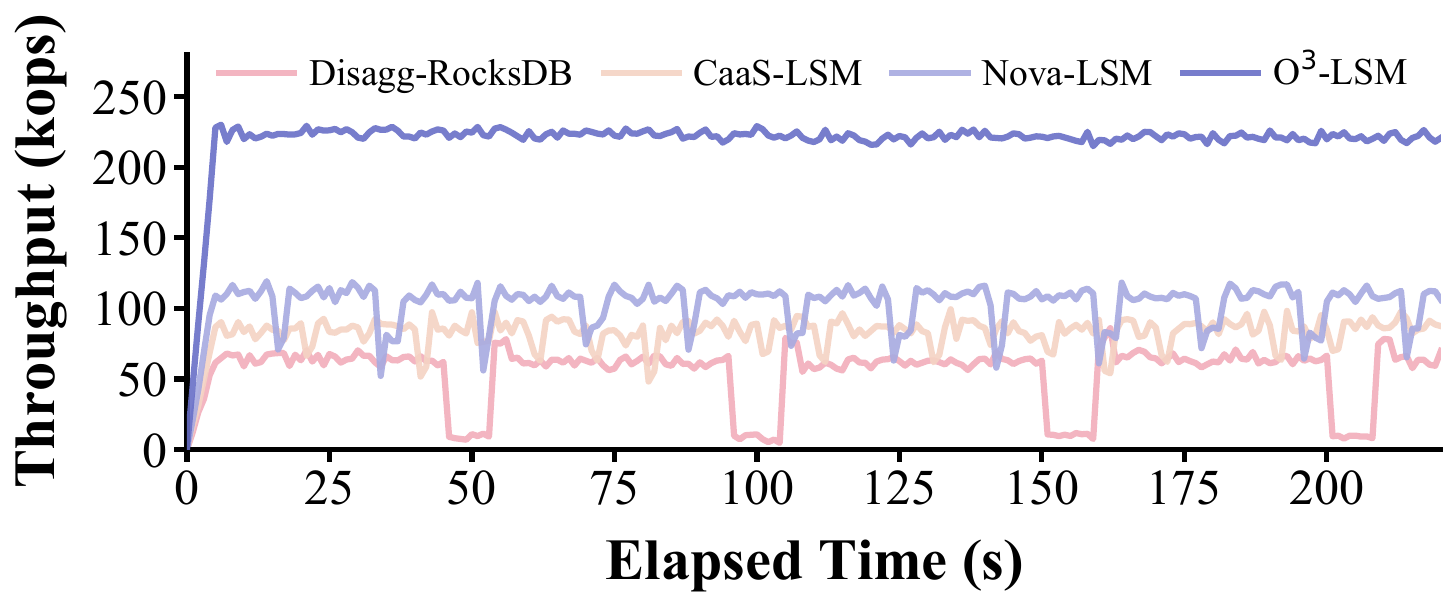}
    \captionsetup{labelformat=parens, labelsep=space}
  \end{minipage}
  \caption{\color{black}Throughput Stability over Time.}
  \label{fig:time-stability}
\end{figure}

\begin{figure}[t]
  \centering
  \begin{minipage}[t]{0.48\textwidth}
    \centering
    \includegraphics[width=\linewidth]{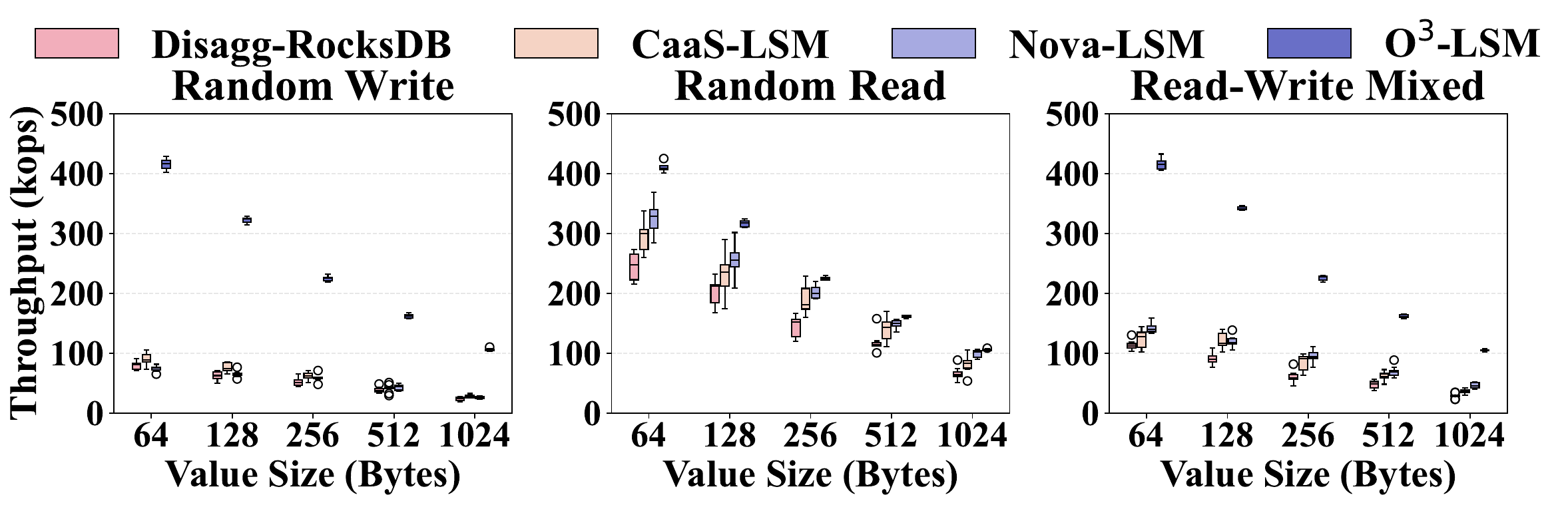}
    \captionsetup{labelformat=parens, labelsep=space}
  \end{minipage}
  \caption{\color{black}Performance Variance and Outlier Analysis.}
  \label{fig:sensitivity-stability}
  \vspace{-5pt}
\end{figure}

\noindent \textcolor{black}{
\textbf{Performance Stability.}
As illustrated in the time-series analysis (\autoref{fig:time-stability}), $O^3$-LSM exhibits high stability compared to other baselines. Disagg-RocksDB and Nova-LSM suffer from frequent throughput dips and wide variance. In contrast, $O^3$-LSM maintains a consistently high and smooth throughput profile. This robustness is further confirmed by the box plots in \autoref{fig:sensitivity-stability}, where $O^3$-LSM displays the most compact distributions and minimal outliers across all workloads and value sizes. This stability stems from replacing coarse-grained flushes with fine-grained shard-level offloading. By partitioning data into small shards and scheduling parallel transfers, O$^3$-LSM avoids the queuing delays and write stalls typical of traditional architectures. This resilience to latency fluctuations makes O$^3$-LSM ideal for latency-sensitive cloud applications.
}

\begin{table}[t]
\centering
\resizebox{\columnwidth}{!}{
\begin{tabular}{@{}lcc@{}}
\toprule
\textbf{Workload} & \textbf{Throughput (kops/s)} & \textbf{P99 Latency ($\mu$s)} \\ \midrule
Random Write & 265.8 (60.1) & 82.4 (169.1) \\
Random Read  & 255.4 (189.1) & 210.1 (440.8) \\
Read-Write Mixed & 272.5 (84.8) & 215.8 (528.0) \\ \bottomrule
\end{tabular}
}
\caption{\color{black}CaaS-LSM w/ and (w/o) $O^3$-LSM Optimizations.}
\label{tab:caas-portability}
\end{table}

\noindent {\color{black} \textbf{Portability of $O^3$-LSM Optimizations.} To evaluate the generalizability of our design, we integrated the $O^3$-LSM memtable and flush offloading mechanisms into CaaS-LSM, a representative compaction offloading LSM-KVS \cite{yu2024caas-lsm}. As shown in \autoref{tab:caas-portability}, the integrated system achieves a Random Write throughput of 265.8~kops/s (a $4.4\times$ improvement over CaaS-LSM's 60.1 kops/s) with a P99 latency of 82.4~$\mu$s (a 51\% reduction from 169.1~$\mu$s). This demonstrates that memtable and flush offloading provide portable benefits. $O^3$-LSM effectively mitigates write stalls and I/O contention within compaction-offloading frameworks. These results underscore $O^3$-LSM's potential as a portable component that maximizes disaggregated LSM-tree efficiency through three-layer offloading.}

\begin{table}[t]
\centering
\small
\setlength{\tabcolsep}{3pt} 
\renewcommand{\arraystretch}{1.1}
\begin{tabular}{@{}lccc@{}}
\toprule
\textbf{Workload} & \textbf{Disagg-RocksDB} & \textbf{CaaS-LSM} & \textbf{O$^3$-LSM} \\
\midrule
SET-only & 86.3 / 391 & 105.9 / 287 & \textbf{381.2 / 177} \\
GET-only & 125.9 / 823 & 159.4 / 751 & \textbf{272.9 / 476} \\
Mixed 10/90 (Get/Set) & 92.5 / 450 & 115.2 / 320 & \textbf{355.4 / 195} \\
Mixed 50/50 (Co-Located) & 102.1 / 610 & 132.8 / 510 & \textbf{315.6 / 310} \\
Mixed 50/50 (Remote) & 72.4 / 1250 & 98.6 / 980 & \textbf{285.4 / 480} \\
Mixed 90/10 (Get/Set) & 118.4 / 780 & 148.6 / 680 & \textbf{285.2 / 410} \\
Pipelined SET (16 Threads) & 185.2 / 1250 & 235.4 / 980 & \textbf{720.5 / 420} \\
\bottomrule
\end{tabular}
\caption{\color{black}Performance Comparison on KVrocks. Metrics are Throughput (kops/s) / P99 Latency (ms).}
\label{tab:kvrocks_comprehensive}
\vspace{-5pt}
\end{table}

\subsection{Analysis of Real-World Application}
We use Kvrocks~\cite{apache_kvrocks} (a Redis-compatible distributed KVS that uses RocksDB as its storage engine) to evaluate the end-to-end performance. Kvrocks integrates three different disaggregated LSM-KVS solutions as its storage engine: Disagg-RocksDB, CaaS-LSM, and O$^3$-LSM. 
\textcolor{black}{We use Redis benchmark and measure throughput and P99 latency for GET-only, SET-only, and mixed GET/SET with 10/90, 50/50, and 90/10 ratios, plus pipelined variants, and we also run both co-located and remote clients. The benchmark issues 40 million queries by 16 threads. As shown in \autoref{tab:kvrocks_comprehensive}, O$^3$-LSM consistently outperforms Disagg-RocksDB and CaaS-LSM across all Redis-benchmark workloads. In SET-only and Mixed 10/90 scenarios, it achieves up to 4.4$\times$ higher throughput and 54\% lower P99 latency. Even under remote client configurations and pipelined workloads, O$^3$-LSM leads CaaS-LSM by 2.9$\times$ and 3.0$\times$ respectively. These gains stem from offloading memtables and parallelizing shard-level flushes, which prevents write stalls and I/O imbalances. These results demonstrate O$^3$-LSM's efficiency in real-world applications.}

\subsection{Scalability}

\begin{figure}
\centering
\begin{subfigure}{.48\columnwidth}
  \centering
  \includegraphics[width=\linewidth]{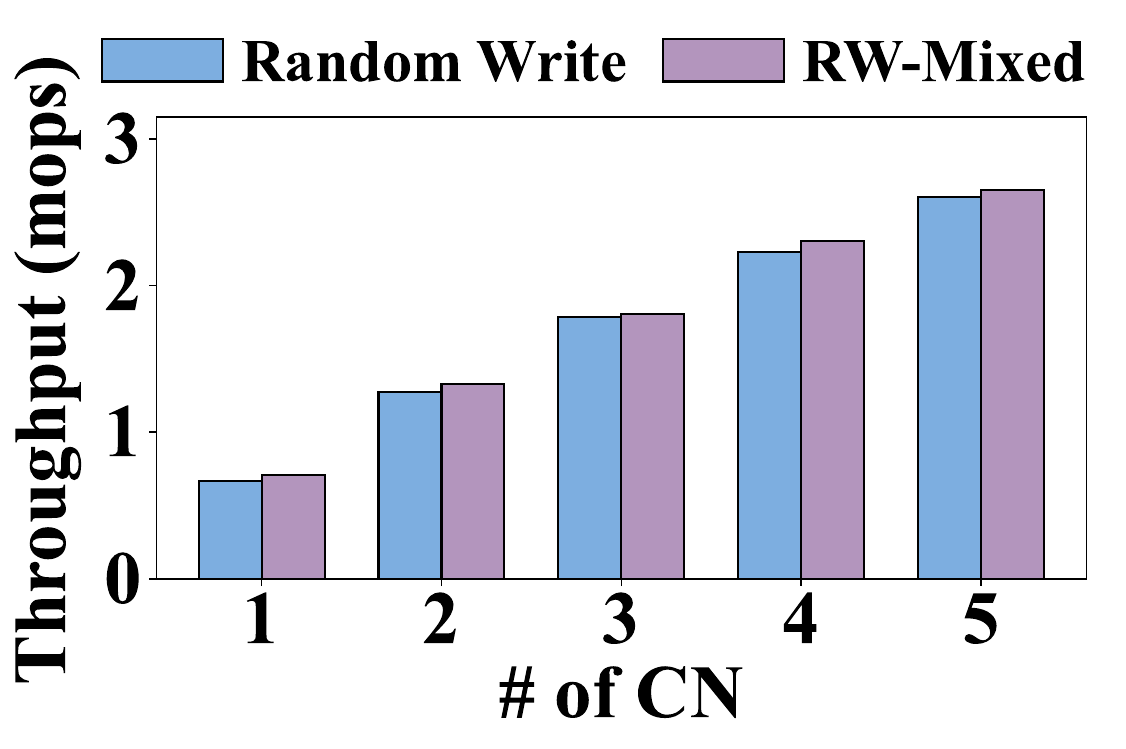}
  \caption{Scale out compute node}
  \label{fig:scale1}
\end{subfigure}%
\begin{subfigure}{.5\columnwidth}
  \centering
  \includegraphics[width=\linewidth]{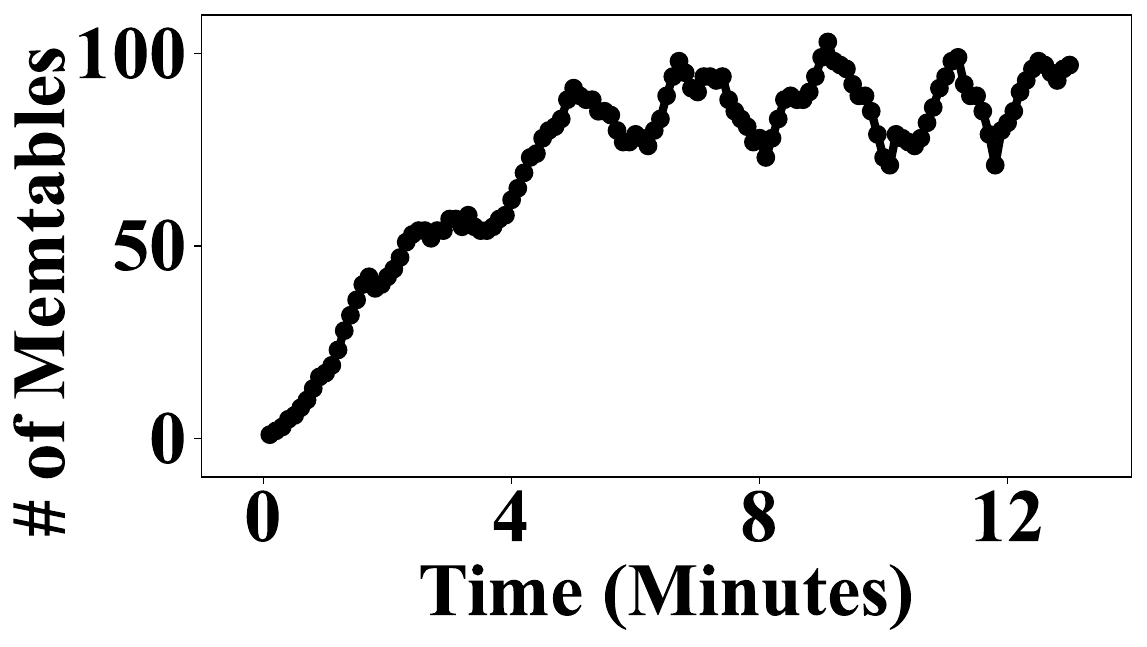}
  \caption{Memtable Counts on DM node}
  \label{fig:scale2}
\end{subfigure}
\caption{Scalability and memory elasticity of O$^3$-LSM.}
\label{fig:scale}
\vspace{-2pt}
\end{figure}

\textbf{Scalability with Multiple CNs.}
We incrementally add CNs under Random Write and Read-Write Mixed workloads to evaluate $O^3$-LSM scalability. Benchmarks were initiated at different time intervals, and we measured the total aggregate throughput across all active CNs. \autoref{fig:scale1} shows the throughput comparison as the number of CNs increases. For both workloads, we observed a consistent increase in throughput, demonstrating good scalability. Adding 4 more CNs improved total throughput by up to 2.9$X$ compared to a single CN. As shown in \autoref{fig:scale2}, with 5 CNs sharing the DM, the total memtable counts demonstrate small fluctuations over an extended period. This stability reflects improved memory utilization and high efficiency, aligning with modern DDC objectives. By balancing flush operations across CNs, $O^3$-LSM prevents simultaneous I/O pressure on any single node, ensuring stable flushes.

\noindent \textcolor{black}{
\textbf{Scalability with Multiple DM Nodes.} We evaluated $O^3$-LSM with multiple DM nodes serving a CN cluster (3 CNs). As shown in \autoref{tab:dm_scalability}, performance gains scale with the number of DM nodes. Upgrading from 1 to 3 DMs boosts the total aggregate throughput by 10.8\% for Random Writes and 11.9\% for RW-Mixed workloads. Each CN independently selects a target DM for offloading and stores the location (Node ID and memory address) within its local memtable metadata. By maintaining this tracking information locally, CNs can concurrently perform tasks (such as memtable offloading and read delegation) across multiple DM nodes. This design parallelizes the offloading overhead and prevents single-node bottlenecks for network I/O or CPU-intensive tasks. Such architectural flexibility allows $O^3$-LSM to linearly expand its memory-tier resources to support increasing concurrent requests and larger working sets in disaggregated environments.
}
\begin{table}[t]
\centering
\small
\renewcommand{\arraystretch}{1.1}
\setlength{\tabcolsep}{8pt}
\begin{tabular}{@{}l cc@{}}
\toprule
\textbf{Configuration} & \textbf{Random Write (kops/s)} & \textbf{RW-Mixed (kops/s)} \\
\midrule
3 CN + 1 DM & 1,786 & 1,805 \\
3 CN + 2 DM & 1,920 & 1,950 \\
3 CN + 3 DM & 1,980 & 2,020 \\
\bottomrule
\end{tabular}
\caption{\color{black}Performance with varying numbers of DM nodes.}
\label{tab:dm_scalability}
\vspace{-5pt}
\end{table}


\noindent \textcolor{black}{
\textbf{Write Stall Mitigation via Shared DM Pool.} 
In the fixed partitioning configuration, each of 5 CNs is assigned a 0.5~GB memory quota at DM (2.5~GB in total). Under this setup, individual CNs frequently exhaust their quotas during write bursts, triggering 840 total stalls and a high P99 latency of 240~ms. Conversely, the shared DM pool configuration utilizes the aggregate 2.5~GB (the same memory budget) capacity to elastically absorb skewed write traffic across the 5 CNs. This collective buffering reduces the total stall count by 92.2\% (from 840 to 65) and slashes P99 latency by over $5\times$, as shown in \autoref{tab:shared_dm_stalls}. By decoupling memory resources from specific compute instances and organizing them as a shared memory pool, $O^3$-LSM allows instances with high write intensity to borrow idle capacity from the pool. This elasticity, paired with parallelized shard flushes, maintains a high aggregate total throughput of 2.62 Mops/s (524 Kops/s per CN), a 41.6\% improvement over the 1.85 Mops/s (370 Kops/s per CN) achieved by the fixed configuration.
}

\vspace{-2pt}
\section{Related Work}
\vspace{-2pt}
\myline{Persistent Disaggregated LSM-KVS} Currently, there are a number of studies focusing on optimizing persistent LSM-KVS on disaggregated storage, including Kemme et al \cite{Ahmad_Kemme_2015}, Hailstorm \cite{Bindschaedler2020hailstorm}, Nova-LSM \cite{Huang_2021}, IS-HBase \cite{cao2022is-hbase}, RocksDB-Cloud \cite{rockscloud}, TerarkDB \cite{terark}, and DisaggreRocksDB \cite{dong2023rocksdb}. Compaction offloading or remote compaction (first proposed in \cite{Ahmad_Kemme_2015}) is mainly used in those studies to reduce the network I/Os between storage and compute nodes. Compared with existing studies of optimizing LSM-KVS for DS, O$^3$-LSM leverages the DM to achieve memory extension and explicit performance improvement. The proposed four major techniques can also be applied to existing techniques (e.g., remote compaction).

\myline{In-memory Key-Value Store for Disaggregated Memory}
Several in-memory key-value store designs have been optimized for DM. Sherman \cite{wang2022sherman} is a B+Tree-based KVS that places all tree nodes in DM while maintaining metadata and index caches locally at CNs, achieving high-performance lookups through designed RDMA access patterns. PolarDB Serverless \cite{cao2021polardb} uses a multi-tenant memory pool over DM to cache data pages, while hash-based systems such as Clover \cite{tsai2020clover}, FUSEE \cite{shen2023fusee}, and Dinomo \cite{lee2022dinomo} store all KV-pairs and indexing structures entirely in DM. 
\textcolor{black}{Recent indexing structures have also explored selective RDMA offloading to balance between one-sided and two-sided RDMA operations. For instance, distributed tree-based indexes~\cite{ziegler2019designing} use this to accelerate traversal, and DEX~\cite{lu2024dex} adopts a similar approach for range indexing on disaggregated memory. $O^3$-LSM uniquely integrates this mechanism within the LSM-tree lifecycle. Unlike these indexing-centric works, our read delegation is co-designed with WAL-based durability and shard-level flush semantics to ensure consistency across the three-layer offloading architecture.}

\myline{LSM-KVS Write Bottleneck} A number of related studies aim to address the write performance bottleneck of LSM-KVS. Some studies \cite{duan2023miodb,kim2022listdb} add an NVM layer to the storage end of LSM-KVS, avoiding serialization overhead and mitigating write amplification by supporting faster flushing. Differently, O$^3$-LSM proposed novel remote flush and cache-enhanced read delegation to successfully address the performance bottleneck caused by slow DM. Various works \cite{Huang_2021,yao2020matrixkv} have different designs for parallelizing $L_0$-L1 compaction. MatrixKV \cite{yao2020matrixkv} maintains a specially partitioned $L_0$ layer on NVM, aiming to fully utilize NVM's high throughput and byte-addressable features. O$^3$-LSM, actively merges the KV range that a single node in the cluster. We aim to offload the memtables on DM and fully utilize the ample compute resources of the shared-DM-attached compute cluster. Other optimizations proposed offloading compaction to other compute resources such as disaggregated storage node CPUs \cite{Ahmad_Kemme_2015,li2021dcompaction,rockscloud, dong2023rocksdb, Huang_2021, cao2022is-hbase}, FPGAs \cite{huang2019xengine} or other compute unit \cite{xu2020luda,ding2023dcomp} to mitigate compaction overhead, or maintaining stability through appropriate compaction strategies and timing \cite{dayan2022spooky,balmau2019silk,yu2023adoc}.

\begin{table}[t]
\centering
\small
\setlength{\tabcolsep}{3pt} 
\renewcommand{\arraystretch}{1.0}
\begin{tabular}{@{}lrrr@{}}
\toprule
\textbf{Setting (5 CNs)} & \textbf{Tput (Mops)} & \textbf{Write Stalls} & \textbf{P99 Latency (ms)} \\
\midrule
Fixed Pool (0.5GB$\times$5) & 1.85 & 840 & 240 \\
Shared Pool (2.5GB) & \textbf{2.62} & \textbf{65} & \textbf{45} \\
\bottomrule
\end{tabular}
\caption{\color{black}Write stall mitigation: Fixed vs. Shared DM Pool.}
\label{tab:shared_dm_stalls}
\vspace{-5pt}
\end{table}
\vspace{-4pt}
\section{Conclusion}

In this paper, we propose $O^3$-LSM, the first LSM-KVS leveraging DM as a write buffer extension for disaggregated storage. We address DM performance challenges via DM-optimized memtables while introducing cache-enhanced read delegation and shard-level flush offloading. Future work includes fine-grained DM memtable management, automated memory ratios, and advanced flush scheduling.

\bibliographystyle{ACM-Reference-Format}
\bibliography{references}

@article{lu2024dex,
  title={DEX: Scalable Range Indexing on Disaggregated Memory},
  author={Lu, Baotong and Huang, Kaisong and Liang, Chieh-Jan Mike and Wang, Tianzheng and Lo, Eric},
  journal={Proceedings of the VLDB Endowment},
  volume={17},
  number={10},
  pages={2603--2616},
  year={2024},
  publisher={VLDB Endowment}
}

@inproceedings{ziegler2019designing,
  title={Designing distributed tree-based index structures for fast rdma-capable networks},
  author={Ziegler, Tobias and Tumkur Vani, Sumukha and Binnig, Carsten and Fonseca, Rodrigo and Kraska, Tim},
  booktitle={Proceedings of the 2019 international conference on management of data},
  pages={741--758},
  year={2019}
}

@article{Ahmad_Kemme_2015,
  title = {Compaction Management in Distributed Key-Value Datastores},
  author = {Ahmad, Muhammad Yousuf and Kemme, Bettina},
  date = {2015-03},
  journaltitle = {Proceedings of the VLDB Endowment},
  pages = {850--861},
  doi = {10.14778/2757807.2757810},
  url = {http://dx.doi.org/10.14778/2757807.2757810},
  langid = {american}
}

@inproceedings{amaro2020fastswap,
  title = {Can far memory improve job throughput?},
  booktitle = {Proceedings of the 15th european conference on computer systems, EuroSys 2020},
  author = {Amaro, Emmanuel and Branner-Augmon, Christopher and Luo, Zhihong and Ousterhout, Amy and Aguilera, Marcos K. and Panda, Aurojit and Ratnasamy, Sylvia and Shenker, Scott},
  date = {2020-04-15},
  series = {Proceedings of the 15th european conference on computer systems, EuroSys 2020},
  publisher = {Association for Computing Machinery, Inc},
  doi = {10.1145/3342195.3387522},
  langid = {English (US)}
}

@inproceedings{angel2020disaggregation,
  title = {Disaggregation and the Application},
  booktitle = {Proceedings of the 12th {{USENIX}} Conference on Hot Topics in Cloud Computing},
  author = {Angel, Sebastian and Nanavati, Mihir and Sen, Siddhartha},
  date = {2020},
  series = {{{HotCloud}}'20},
  publisher = {USENIX Association},
  location = {USA},
  articleno = {15},
  pagetotal = {1}
}

@misc{apache_kvrocks,
  title = {Kvrocks},
  author = {{Apache}},
  url = {https://github.com/apache/incubator-kvrocks}
}

@inproceedings{balmau2019silk,
  title = {{{SILK}}: {{Preventing}} Latency Spikes in {{Log-Structured}} Merge {{Key-Value}} Stores},
  booktitle = {2019 {{USENIX}} Annual Technical Conference ({{USENIX ATC}} 19)},
  author = {Balmau, Oana and Dinu, Florin and Zwaenepoel, Willy and Gupta, Karan and Chandhiramoorthi, Ravishankar and Didona, Diego},
  date = {2019-07},
  pages = {753--766},
  publisher = {USENIX Association},
  location = {Renton, WA},
  url = {https://www.usenix.org/conference/atc19/presentation/balmau},
  isbn = {978-1-939133-03-8}
}

@inproceedings{Bindschaedler2020hailstorm,
  title = {Hailstorm: {{Disaggregated}} Compute and Storage for Distributed {{LSM-based}} Databases},
  booktitle = {Proceedings of the Twenty-Fifth International Conference on Architectural Support for Programming Languages and Operating Systems},
  author = {Bindschaedler, Laurent and Goel, Ashvin and Zwaenepoel, Willy},
  date = {2020},
  series = {Asplos '20},
  pages = {301--316},
  publisher = {Association for Computing Machinery},
  address = {New York, NY, USA},
  location = {Lausanne, Switzerland},
  doi = {10.1145/3373376.3378504},
  url = {https://doi.org/10.1145/3373376.3378504},
  isbn = {978-1-4503-7102-5},
  pagetotal = {16}
}

@inproceedings{cao2020characterizing,
  title = {Characterizing, Modeling, and Benchmarking {{RocksDB}} Key-Value Workloads at Facebook},
  booktitle = {18th {{USENIX}} Conference on File and Storage Technologies ({{FAST}} 20)},
  author = {Cao, Zhichao and Dong, Siying and Vemuri, Sagar and Du, David HC},
  date = {2020},
  pages = {209--223}
}

@inproceedings{cao2021polardb,
  title = {{{PolarDB}} Serverless: A Cloud Native Database for Disaggregated Data Centers},
  booktitle = {Proceedings of the 2021 International Conference on Management of Data},
  author = {Cao, Wei and Zhang, Yingqiang and Yang, Xinjun and Li, Feifei and Wang, Sheng and Hu, Qingda and Cheng, Xuntao and Chen, Zongzhi and Liu, Zhenjun and Fang, Jing and Wang, Bo and Wang, Yuhui and Sun, Haiqing and Yang, Ze and Cheng, Zhushi and Chen, Sen and Wu, Jian and Hu, Wei and Zhao, Jianwei and Gao, Yusong and Cai, Songlu and Zhang, Yunyang and Tong, Jiawang},
  date = {2021},
  series = {Sigmod '21},
  pages = {2477--2489},
  publisher = {Association for Computing Machinery},
  address = {New York, NY, USA},
  location = {Virtual Event, China},
  doi = {10.1145/3448016.3457560},
  url = {https://doi.org/10.1145/3448016.3457560},
  isbn = {978-1-4503-8343-1},
  pagetotal = {13}
}

@article{cao2022is-hbase,
  title = {{{IS-hbase}}: {{An}} in-Storage Computing Optimized Hbase with {{I}}/{{O}} Offloading and Self-Adaptive Caching in Compute-Storage Disaggregated Infrastructure},
  author = {Cao, Zhichao and Dong, Huibing and Wei, Yixun and Liu, Shiyong and Du, David H. C.},
  date = {2022-04},
  journaltitle = {ACM Transactions on Storage},
  shortjournal = {ACM Trans. Storage},
  volume = {18},
  number = {2},
  publisher = {Association for Computing Machinery},
  location = {New York, NY, USA},
  issn = {1553-3077},
  doi = {10.1145/3488368},
  url = {https://doi.org/10.1145/3488368},
  articleno = {15},
  issue_date = {May 2022},
  pagetotal = {42}
}

@inproceedings{cloudlab,
  title = {The Design and Operation of {{CloudLab}}},
  booktitle = {2019 {{USENIX}} Annual Technical Conference ({{USENIX ATC}} 19)},
  author = {Duplyakin, Dmitry and Ricci, Robert and Maricq, Aleksander and Wong, Gary and Duerig, Jonathon and Eide, Eric and Stoller, Leigh and Hibler, Mike and Johnson, David and Webb, Kirk and Akella, Aditya and Wang, Kuangching and Ricart, Glenn and Landweber, Larry and Elliott, Chip and Zink, Michael and Cecchet, Emmanuel and Kar, Snigdhaswin and Mishra, Prabodh},
  date = {2019-07},
  pages = {1--14},
  publisher = {USENIX Association},
  location = {Renton, WA},
  url = {https://www.usenix.org/conference/atc19/presentation/duplyakin},
  isbn = {978-1-939133-03-8}
}

@online{cxl,
  title = {{{CXL}} Consortium},
  url = {https://www.computeexpresslink.org},
  author = {{Compute Express Link (CXL) Consortium}},
  year = {2019}
}

@misc{cxl-3.0,
  title = {{{CXL}} 3.0 White Paper},
  author = {{Compute Express Link (CXL) Consortium}},
  year = {Accessed: November, 2022},
  url = {https://www.computeexpresslink.org/_files/ugd/0c1418_a8713008916044ae9604405d10a7773b.pdf}
}

@article{dayan2022spooky,
  title = {Spooky: Granulating {{LSM-tree}} Compactions Correctly},
  author = {Dayan, Niv and Weiss, Tamar and Dashevsky, Shmuel and Pan, Michael and Bortnikov, Edward and Twitto, Moshe},
  date = {2022-07},
  journaltitle = {Proc. VLDB Endow.},
  volume = {15},
  number = {11},
  pages = {3071--3084},
  publisher = {VLDB Endowment},
  issn = {2150-8097},
  doi = {10.14778/3551793.3551853},
  url = {https://doi.org/10.14778/3551793.3551853},
  issue_date = {July 2022},
  pagetotal = {14}
}

@inproceedings{ding2023dcomp,
  title = {{{DComp}}: {{Efficient}} Offload of {{LSM-tree}} Compaction with Data Processing Units},
  booktitle = {Proceedings of the 52nd International Conference on Parallel Processing},
  author = {Ding, Chen and Zhou, Jian and Wan, Jiguang and Xiong, Yiqin and Li, Sicen and Chen, Shuning and Liu, Hanyang and Tang, Liu and Zhan, Ling and Lu, Kai and Xu, Peng},
  date = {2023},
  series = {Icpp '23},
  pages = {233--243},
  publisher = {Association for Computing Machinery},
  address = {New York, NY, USA},
  location = {¡conf-loc¿, ¡city¿Salt Lake City¡/city¿, ¡state¿UT¡/state¿, ¡country¿USA¡/country¿, ¡/conf-loc¿},
  doi = {10.1145/3605573.3605633},
  url = {https://doi.org/10.1145/3605573.3605633},
  isbn = {979-8-4007-0843-5},
  pagetotal = {11}
}

@online{dm,
  title = {Disaggregated Memory},
  author = {{IBM}},
  date = {2018},
  url = {https://www.ibm.com/blogs/research/2018/01/advancing-cloud-memory-disaggregation/}
}

@article{dong2023rocksdb,
  title = {Disaggregating {{RocksDB}}: A Production Experience},
  author = {Dong, Siying and P, Shiva Shankar and Pan, Satadru and Ananthabhotla, Anand and Ekambaram, Dhanabal and Sharma, Abhinav and Dayal, Shobhit and Parikh, Nishant Vinaybhai and Jin, Yanqin and Kim, Albert and Patil, Sushil and Zhuang, Jay and Dunster, Sam and Mahajan, Akanksha and Chelluri, Anirudh and Datye, Chaitanya and Santana, Lucas Vasconcelos and Garg, Nitin and Gawde, Omkar},
  date = {2023-06},
  journaltitle = {Proc. ACM Manag. Data},
  volume = {1},
  number = {2},
  publisher = {Association for Computing Machinery},
  location = {New York, NY, USA},
  doi = {10.1145/3589772},
  url = {https://doi.org/10.1145/3589772},
  articleno = {192},
  issue_date = {June 2023},
  pagetotal = {24}
}

@inproceedings{duan2023miodb,
  title = {Revisiting Log-Structured Merging for {{KV}} Stores in Hybrid Memory Systems},
  booktitle = {Proceedings of the 28th {{ACM}} International Conference on Architectural Support for Programming Languages and Operating Systems, Volume 2},
  author = {Duan, Zhuohui and Yao, Jiabo and Liu, Haikun and Liao, Xiaofei and Jin, Hai and Zhang, Yu},
  date = {2023},
  series = {Asplos 2023},
  pages = {674--687},
  publisher = {Association for Computing Machinery},
  address = {New York, NY, USA},
  location = {Vancouver, BC, Canada},
  doi = {10.1145/3575693.3575715},
  url = {https://doi.org/10.1145/3575693.3575715},
  isbn = {978-1-4503-9916-6},
  pagetotal = {14}
}

@inproceedings{gao2016network,
  title = {Network Requirements for Resource Disaggregation},
  booktitle = {12th {{USENIX}} Symposium on Operating Systems Design and Implementation ({{OSDI}} 16)},
  author = {Gao, Peter X. and Narayan, Akshay and Karandikar, Sagar and Carreira, Joao and Han, Sangjin and Agarwal, Rachit and Ratnasamy, Sylvia and Shenker, Scott},
  date = {2016-11},
  pages = {249--264},
  publisher = {USENIX Association},
  location = {Savannah, GA},
  url = {https://www.usenix.org/conference/osdi16/technical-sessions/presentation/gao},
  isbn = {978-1-931971-33-1}
}

@inproceedings{ghemawat2003google,
  title = {The {{Google}} File System},
  booktitle = {Proceedings of the Nineteenth {{ACM}} Symposium on {{Operating}} Systems Principles},
  author = {Ghemawat, Sanjay and Gobioff, Howard and Leung, Shun-Tak},
  date = {2003},
  pages = {29--43}
}

@inproceedings{hdfs,
  title = {The Hadoop Distributed File System},
  booktitle = {2010 {{IEEE}} 26th Symposium on Mass Storage Systems and Technologies ({{MSST}} 10)},
  author = {Shvachko, Konstantin and Kuang, Hairong and Radia, Sanjay and Chansler, Robert},
  date = {2010},
  pages = {1--10},
  publisher = {IEEE}
}

@inproceedings{hpcc12rdma,
  title = {A Performance Study to Guide {{RDMA}} Programming Decisions},
  booktitle = {Proceedings of the 2012 {{IEEE}} 14th International Conference on High Performance Computing and Communication \& 2012 {{IEEE}} 9th International Conference on Embedded Software and Systems},
  author = {MacArthur, Patrick and Russell, Robert D.},
  date = {2012},
  series = {Hpcc '12},
  pages = {778--785},
  publisher = {IEEE Computer Society},
  location = {USA},
  doi = {10.1109/HPCC.2012.110},
  url = {https://doi.org/10.1109/HPCC.2012.110},
  isbn = {978-0-7695-4749-7},
  pagetotal = {8}
}

@inproceedings{Huang_2021,
  title = {Nova-{{LSM}}: A Distributed, Component-Based {{LSM-tree}} Key-Value Store},
  booktitle = {Proceedings of the 2021 International Conference on Management of Data},
  author = {Huang, Haoyu and Ghandeharizadeh, Shahram},
  date = {2021-06},
  publisher = {ACM},
  doi = {10.1145/3448016.3457297},
  url = {https://doi.org/10.1145%2F3448016.3457297}
}

@inproceedings{huang2019xengine,
  title = {X-Engine: {{An}} Optimized Storage Engine for Large-Scale e-Commerce Transaction Processing},
  booktitle = {Proceedings of the 2019 International Conference on Management of Data},
  author = {Huang, Gui and Cheng, Xuntao and Wang, Jianying and Wang, Yujie and He, Dengcheng and Zhang, Tieying and Li, Feifei and Wang, Sheng and Cao, Wei and Li, Qiang},
  date = {2019},
  series = {Sigmod '19},
  pages = {651--665},
  publisher = {Association for Computing Machinery},
  address = {New York, NY, USA},
  location = {Amsterdam, Netherlands},
  doi = {10.1145/3299869.3314041},
  url = {https://doi.org/10.1145/3299869.3314041},
  isbn = {978-1-4503-5643-5},
  pagetotal = {15}
}

@online{inifiniband,
  title = {Infiniband},
  author = {{Oracle}},
  date = {2012},
  url = {https://www.oracle.com/technetwork/server-storage/networking/documentation/o12-020-1653901.pdf}
}

@inproceedings{kim2022listdb,
  title = {{{ListDB}}: {{Union}} of {{Write-Ahead}} Logs and Persistent {{SkipLists}} for Incremental Checkpointing on Persistent Memory},
  booktitle = {16th {{USENIX}} Symposium on Operating Systems Design and Implementation ({{OSDI}} 22)},
  author = {Kim, Wonbae and Park, Chanyeol and Kim, Dongui and Park, Hyeongjun and Choi, Young-ri and Sussman, Alan and Nam, Beomseok},
  date = {2022-07},
  pages = {161--177},
  publisher = {USENIX Association},
  location = {Carlsbad, CA},
  url = {https://www.usenix.org/conference/osdi22/presentation/kim},
  isbn = {978-1-939133-28-1}
}

@article{lee2022dinomo,
  title = {{{DINOMO}}: {{An}} Elastic, Scalable, High-Performance Key-Value Store for Disaggregated Persistent Memory},
  author = {Lee, Sekwon and Ponnapalli, Soujanya and Singhal, Sharad and Aguilera, Marcos K. and Keeton, Kimberly and Chidambaram, Vijay},
  date = {2022-09},
  journaltitle = {Proc. VLDB Endow.},
  volume = {15},
  number = {13},
  pages = {4023--4037},
  publisher = {VLDB Endowment},
  issn = {2150-8097},
  doi = {10.14778/3565838.3565854},
  url = {https://doi.org/10.14778/3565838.3565854},
  issue_date = {September 2022},
  pagetotal = {15}
}

@article{leveldb,
  title = {{{LevelDB}}},
  author = {Ghemawat, Sanjay and Dean, Jeff},
  date = {2011},
  journaltitle = {URL: https://github. com/google/leveldb,\% 20http://leveldb. org}
}

@inproceedings{li2021dcompaction,
  title = {Elastic and Stable Compaction for {{LSM-tree}}: A {{FaaS-based}} Approach on {{TerarkDB}}},
  booktitle = {Proceedings of the 30th {{ACM}} International Conference on Information \& Knowledge Management},
  author = {Li, Jianchuan and Jin, Peiquan and Lin, Yuanjin and Zhao, Ming and Wang, Yi and Guo, Kuankuan},
  date = {2021},
  series = {Cikm '21},
  pages = {3906--3915},
  publisher = {Association for Computing Machinery},
  address = {New York, NY, USA},
  location = {Virtual Event, Queensland, Australia},
  doi = {10.1145/3459637.3481913},
  url = {https://doi.org/10.1145/3459637.3481913},
  isbn = {978-1-4503-8446-9},
  pagetotal = {10}
}

@article{li2022dm,
  title = {First-Generation Memory Disaggregation for Cloud Platforms},
  author = {Li, Huaicheng and Berger, Daniel and Novakovic, Stanko and Hsu, Lisa and Ernst, Dan and Zardoshti, Pantea and Shah, Monish and Agarwal, Ishwar and Hill, Mark and Fontoura, Marcus and Bianchini, Ricardo},
  date = {2022-03}
}

@article{lin2020ddc,
  title = {Disaggregated Data Centers: {{Challenges}} and Trade-Offs},
  author = {Lin, Rui and Cheng, Yuxin and Andrade, Marilet De and Wosinska, Lena and Chen, Jiajia},
  date = {2020},
  journaltitle = {IEEE Communications Magazine},
  volume = {58},
  number = {2},
  pages = {20--26},
  doi = {10.1109/MCOM.001.1900612}
}

@inproceedings{liu2021consistent,
  title = {Consistent Rdma-Friendly Hashing on Remote Persistent Memory},
  booktitle = {2021 {{IEEE}} 39th International Conference on Computer Design ({{ICCD}})},
  author = {Liu, Xinxin and Hua, Yu and Bai, Rong},
  date = {2021},
  pages = {174--177},
  publisher = {IEEE},
  doi = {10.1109/ICCD53106.2021.00037}
}

@article{nag2023cxl,
  title = {{{CXL}} (Compute Express Link) Technology},
  author = {Nag, Santhosh Nagaraj},
  date = {2023},
  journaltitle = {Journal of Computer and Communications},
  volume = {11},
  number = {6},
  pages = {94--102},
  publisher = {Scientific Research Publishing}
}

@book{O3-lsm,
  title = {{{https://github.com/argetterxx/O3-LSM}}},
  date = {2024},
  author = {Qi Lin, Gangqi Huang}
}

@online{rockscloud,
  title = {Rockscloud},
  author = {{rockset}},
  date = {2020},
  url = {https://github.com/rockset/rocksdb-cloud}
}

@inproceedings{rocksdb,
  title = {Optimizing Space Amplification in {{RocksDB}}},
  booktitle = {{{CIDR}}},
  author = {Dong, Siying and Callaghan, Mark and Galanis, Leonidas and Borthakur, Dhruba and Savor, Tony and Strum, Michael},
  date = {2017},
  volume = {3},
  pages = {3}
}

@inproceedings{shen2023fusee,
  title = {{{FUSEE}}: A Fully Memory-Disaggregated Key-Value Store},
  booktitle = {Proceedings of the 21st {{USENIX}} Conference on File and Storage Technologies},
  author = {Shen, Jiacheng and Zuo, Pengfei and Luo, Xuchuan and Yang, Tianyi and Su, Yuxin and Zhou, Yangfan and Lyu, Michael R.},
  date = {2023},
  series = {{{FAST}}'23},
  publisher = {USENIX Association},
  address = {USA},
  location = {Santa Clara, CA, USA},
  articleno = {6},
  isbn = {978-1-939133-32-8},
  pagetotal = {17}
}

@inproceedings{shoal2019,
  title = {Shoal: A Network Architecture for Disaggregated Racks},
  booktitle = {16th {{USENIX}} Symposium on Networked Systems Design and Implementation ({{NSDI}} 19)},
  author = {Shrivastav, Vishal and Valadarsky, Asaf and Ballani, Hitesh and Costa, Paolo and Lee, Ki Suh and Wang, Han and Agarwal, Rachit and Weatherspoon, Hakim},
  date = {2019-02},
  pages = {255--270},
  publisher = {USENIX Association},
  location = {Boston, MA},
  url = {https://www.usenix.org/conference/nsdi19/presentation/shrivastav},
  isbn = {978-1-931971-49-2}
}

@inproceedings{tectonic,
  title = {Facebook’s Tectonic Filesystem: {{Efficiency}} from Exascale},
  booktitle = {19th {{USENIX}} Conference on File and Storage Technologies ({{FAST}} 21)},
  author = {Pan, Satadru and Stavrinos, Theano and Zhang, Yunqiao and Sikaria, Atul and Zakharov, Pavel and Sharma, Abhinav and P, Shiva Shankar and Shuey, Mike and Wareing, Richard and Gangapuram, Monika and Cao, Guanglei and Preseau, Christian and Singh, Pratap and Patiejunas, Kestutis and family=Tipton, given=JR, given-i=JR and Katz-Bassett, Ethan and Lloyd, Wyatt},
  date = {2021-02},
  pages = {217--231},
  publisher = {USENIX Association},
  url = {https://www.usenix.org/conference/fast21/presentation/pan},
  isbn = {978-1-939133-20-5}
}

@online{terark,
  title = {Terark},
  author = {{bytedance}},
  date = {2019},
  url = {https://github.com/bytedance/terarkdb}
}

@inproceedings{tsai2020clover,
  title = {Disaggregating Persistent Memory and Controlling Them Remotely: {{An}} Exploration of Passive Disaggregated Key-Value Stores},
  booktitle = {Proceedings of the 2020 {{USENIX}} Conference on Usenix Annual Technical Conference},
  author = {Tsai, Shin-Yeh and Shan, Yizhou and Zhang, Yiying},
  date = {2020},
  series = {Usenix Atc'20},
  publisher = {USENIX Association},
  location = {USA},
  articleno = {3},
  isbn = {978-1-939133-14-4},
  pagetotal = {16}
}

@inproceedings{wang2022sherman,
  title = {Sherman: A Write-Optimized Distributed {{B}}+tree Index on Disaggregated Memory},
  booktitle = {Proceedings of the 2022 International Conference on Management of Data},
  author = {Wang, Qing and Lu, Youyou and Shu, Jiwu},
  date = {2022},
  series = {Sigmod '22},
  pages = {1033--1048},
  publisher = {Association for Computing Machinery},
  address = {New York, NY, USA},
  location = {Philadelphia, PA, USA},
  doi = {10.1145/3514221.3517824},
  url = {https://doi.org/10.1145/3514221.3517824},
  isbn = {978-1-4503-9249-5},
  pagetotal = {16}
}

@inproceedings{wang2023disaggregated,
  title = {Disaggregated Database Systems},
  booktitle = {Companion of the 2023 International Conference on Management of Data},
  author = {Wang, Jianguo and Zhang, Qizhen},
  date = {2023},
  pages = {37--44}
}

@inproceedings{wang2023dlsm,
  title = {{{dLSM}}: {{An LSM-based}} Index for Memory Disaggregation},
  booktitle = {2023 {{IEEE}} 39th International Conference on Data Engineering ({{ICDE}})},
  author = {Wang, Ruihong and Wang, Jianguo and Kadam, Prishita and Özsu, M Tamer and Aref, Walid G},
  date = {2023},
  pages = {2835--2849},
  publisher = {IEEE}
}

@misc{xu2020luda,
  title = {{{LUDA}}: {{Boost LSM}} Key Value Store Compactions with Gpus},
  author = {Xu, Peng and Wan, Jiguang and Huang, Ping and Yang, Xiaogang and Tang, Chenlei and Wu, Fei and Xie, Changsheng},
  date = {2020},
  eprint = {2004.03054},
  eprinttype = {arXiv},
  eprintclass = {cs.DC}
}

@inproceedings{yao2020matrixkv,
  title = {{{MatrixKV}}: {{Reducing}} Write Stalls and Write Amplification in {{LSM-tree}} Based {{KV}} Stores with a Matrix Container in {{NVM}}},
  booktitle = {Proceedings of the 2020 {{USENIX}} Conference on Usenix Annual Technical Conference},
  author = {Yao, Ting and Zhang, Yiwen and Wan, Jiguang and Cui, Qiu and Tang, Liu and Jiang, Hong and Xie, Changsheng and He, Xubin},
  date = {2020},
  series = {Usenix Atc'20},
  publisher = {USENIX Association},
  location = {USA},
  articleno = {2},
  isbn = {978-1-939133-14-4},
  pagetotal = {15}
}

@inproceedings{ycsb,
  title = {Benchmarking Cloud Serving Systems with {{YCSB}}},
  booktitle = {Proceedings of the 1st {{ACM}} Symposium on Cloud Computing},
  author = {Cooper, Brian F and Silberstein, Adam and Tam, Erwin and Ramakrishnan, Raghu and Sears, Russell},
  date = {2010},
  pages = {143--154},
  publisher = {ACM}
}

@inproceedings{yoo2022read,
  title = {A Read Performance Analysis with Storage Hierarchy in Modern {{KVS}}: A {{RocksDB}} Case},
  booktitle = {2022 {{IEEE}} 11th Non-Volatile Memory Systems and Applications Symposium ({{NVMSA}})},
  author = {Yoo, Seehwan and Shin, Hojin and Lee, Sunghyun and Choi, Jongmoo},
  date = {2022},
  pages = {45--50},
  doi = {10.1109/NVMSA56066.2022.00017}
}

@inproceedings{yu2023adoc,
  title = {{{ADOC}}: {{Automatically}} Harmonizing Dataflow between Components in {{Log-Structured Key-Value}} Stores for Improved Performance},
  booktitle = {21st {{USENIX}} Conference on File and Storage Technologies ({{FAST}} 23)},
  author = {Yu, Jinghuan and Noh, Sam H. and Choi, Young-ri and Xue, Chun Jason},
  date = {2023-02},
  pages = {65--80},
  publisher = {USENIX Association},
  location = {Santa Clara, CA},
  url = {https://www.usenix.org/conference/fast23/presentation/yu},
  isbn = {978-1-939133-32-8}
}

@article{yu2024caas-lsm,
  title = {{{CaaS-LSM}}: {{Compaction-as-a-service}} for {{LSM-based}} Key-Value Stores in Storage Disaggregated Infrastructure.},
  author = {Yu, Qiaolin and Guo, Chang and Zhuang, Jay and Thakkar, Viraj and Wang, Jianguo and Cao, Zhichao},
  date = {2024},
  journaltitle = {Proceedings of the ACM on Management of Data},
  volume = {2},
  number = {3},
  pages = {1--26},
  publisher = {ACM New York, NY, USA}
}

@article{zhang2020ddc,
  title = {Understanding the Effect of Data Center Resource Disaggregation on Production {{DBMSs}}},
  author = {Zhang, Qizhen and Cai, Yifan and Chen, Xinyi and Angel, Sebastian and Chen, Ang and Liu, Vincent and Loo, Boon Thau},
  date = {2020-05},
  journaltitle = {Proc. VLDB Endow.},
  volume = {13},
  number = {9},
  pages = {1568--1581},
  publisher = {VLDB Endowment},
  issn = {2150-8097},
  doi = {10.14778/3397230.3397249},
  url = {https://doi.org/10.14778/3397230.3397249},
  issue_date = {May 2020},
  pagetotal = {14}
}

@inproceedings{zhang2020rethinking,
  title = {Rethinking Data Management Systems for Disaggregated Data Centers},
  booktitle = {Conference on Innovative Data Systems Research},
  author = {Zhang, Qizhen and Cai, Yifan and Angel, Sebastian and Chen, Ang and Liu, Vincent and Loo, Boon Thau},
  date = {2020}
}

@online{zippydb,
  title = {{{ZippyDB}}},
  author = {{Facebook}},
  date = {2021},
  url = {https://engineering.fb.com/2021/08/06/core-infra/zippydb/}
}










\end{document}